\documentclass[11pt]{article}

\usepackage[utf8]{inputenc} 
\usepackage[english]{babel}
\usepackage[T1]{fontenc}
\usepackage[margin=23mm, vmargin=32mm]{geometry}
\usepackage{caption}
\usepackage{tikz-cd}
\usepackage{amsmath,amsfonts,amssymb,amsthm}
\usepackage{authblk}
\usepackage{url}
\usepackage{setspace}
\usepackage{cancel,verbatim,slashed}
\usepackage{xcolor,mdframed,graphicx,color}
\usepackage{mathtools}
\usepackage{multirow,stmaryrd}
\usepackage{array}
\usepackage{makecell}
\usepackage{upgreek}
\usepackage{mathrsfs}
\usepackage{braket}
\usepackage{pict2e}
\usepackage{stackengine}
\usepackage{multirow}
\usepackage{multicol}
\usepackage[titles]{tocloft}
\usepackage{newpxtext}
\usepackage{newpxmath}

\newcommand{\BU}{\mathbf{B}\mathrm{U}(1)_\mathrm{conn}}
\newcommand{\Coo}{\mathcal{C}^{\infty}}
\newcommand{\Aut}{\mathbf{Aut}}
\newcommand{\Mfd}{\mathsf{Mfd}}

\newcommand{\B}{\mathbf{B}}
\newcommand{\di}{\mathrm{d}}

\newcommand{\CE}{\mathrm{CE}}

\newcommand{\op}{\mathrm{op}}
\newcommand{\Hom}{\mathrm{Hom}}

\newcommand{\sSet}{\mathsf{sSet}}
\newcommand{\topos}{\mathbf{H}}

\newcommand{\bbR}{\mathbb{R}}

\newcommand{\bbT}{\mathbb{T}}

\newcommand{\dR}{\mathrm{dR}}
\newcommand{\BV}{\mathrm{BV}}

\DeclareRobustCommand\longtwoheadrightarrow
    {\relbar\joinrel\twoheadrightarrow}

\tikzcdset{%
    triple line/.code={\tikzset{%
        double equal sign distance, 
        double=\pgfkeysvalueof{/tikz/commutative diagrams/background color}}},
    quadruple line/.code={\tikzset{%
        double equal sign distance, 
        double=\pgfkeysvalueof{/tikz/commutative diagrams/background color}}},
    Rrightarrow/.code={\tikzcdset{triple line}\pgfsetarrows{tikzcd implies cap-tikzcd implies}},
    RRightarrow/.code={\tikzcdset{quadruple line}\pgfsetarrows{tikzcd implies cap-tikzcd implies}}
}    
\newcommand*{\tarrow}[2][]{\arrow[Rrightarrow, #1]{#2}\arrow[dash, shorten >= 0.5pt, #1]{#2}}

\makeatletter
\providecommand{\leftsquigarrow}{%
  \mathrel{\mathpalette\reflect@squig\relax}%
}
\newcommand{\reflect@squig}[2]{%
  \reflectbox{$\m@th#1\rightsquigarrow$}%
}
\makeatother

\usepackage{ textcomp }

\usepackage[style=alphabetic, sorting=nyt, backend=biber, doi=false, isbn=false]{biblatex}
\usepackage[hypertexnames=true]{hyperref}
\newcommand*{\bibtitle}{References}

\renewcommand*{\bibfont}{\raggedright\small}

\usepackage{cleveref}

\title{\bf \vspace{-1.0cm}
Higher geometry in physics}
\author{{\sc Luigi Alfonsi}\vspace{-2mm}
}
\affil{\em\normalsize Department of Physics, Astronomy and Mathematics,\\\em University of Hertfordshire, Hatfield AL10 9AB, UK\\\vspace{2mm}\tt \href{mailto:l.alfonsi@herts.ac.uk}{l.alfonsi@herts.ac.uk} \vspace{-1mm}
}
\date{\small December 12, 2023}\vspace{-5mm}

\begin{document}

\maketitle
\abstract{
\noindent 
This survey article is an invited contribution to the Encyclopedia of Mathematical Physics, $2$nd edition. We provide an accessible overview on relevant applications of higher and derived geometry to theoretical physics, including higher gauge theory, higher geometric quantization and Batalin-Vilkovisky formalism.

\vspace{2mm}

\vspace{2mm}
}

\tableofcontents

\section*{Introduction}
\addcontentsline{toc}{section}{Introduction}
\setlength{\parindent}{0pt}
\setlength{\parskip}{0.5em}

\textit{Higher geometry} is a generalization of ordinary geometry rooted in higher category theory, at which core lies the concept of $\infty$-category. 
While traditional categories are made of objects and composable morphisms, $\infty$-categories introduce a hierarchy of relationships beyond ordinary morphisms. Informally speaking, in an $\infty$-category, one can consider morphisms between morphisms, morphisms between morphisms between morphisms, and so on. 

Higher geometric spaces are constructed by probing into them with ordinary spaces, in an $\infty$-category. More precisely, they are constructed as stacks, i.e. $\infty$-categorical sheaves.
Within the umbrella of higher geometry, we also have (higher) derived geometry, where the probing spaces are also categorified.

Despite the complicated mathematical machinery needed, from a physical perspective, these generalizations of ordinary geometry are very natural: they lead to a geometry which naturally encompasses gauge fields and Feynman diagrams (see table \ref{tab:geometries}).

\begin{table}[h!]\begin{center}
\begin{center}
 \begin{center}
\begin{tabular}{||c | c c||} 
 \hline
   & Scalar fields & Gauge fields \\ [0.5ex] 
 \hline
 Off-shell fields & Ordinary geometry & Higher geometry \\ 
 Feynman diagrams & Derived geometry  & Higher derived geometry \\ [1ex] 
 \hline
\end{tabular}
\end{center}
\end{center}
\caption{\label{tab:geometries}Geometric frameworks within the umbrella of higher geometry, and physical meaning.}\vspace{-0.5cm}
\end{center}\end{table}

Higher geometry can be seen as differential geometry where equality has been replaced by the weaker notion of equivalence. In physics, it is not typically relevant to question whether two field configurations are equal, but rather whether the they are gauge-equivalent. For some fields, gauge-equivalences can in turn be related by gauge-of-gauge-equivalences, and so on.
\vspace{-0.2cm}
\begin{equation*}
{\begin{tikzcd}[row sep=scriptsize, column sep=12ex]
    \phi \arrow[r, ""]
    & \phi'
\end{tikzcd}}, \quad\;
\begin{tikzcd}[row sep=scriptsize, column sep=14ex]
    \phi \arrow[r, bend left=60, ""{name=U, below}, "\,"]
    \arrow[r, bend right=60, ""{name=D}, ""']
    & \phi'
    \arrow[Rightarrow, from=U, to=D, "", end anchor={[xshift=-0.18ex]}]
\end{tikzcd}, \quad\;
\begin{tikzcd}[row sep=scriptsize, column sep=16ex]
    \phi \arrow[r, bend left=60, ""{name=U, below}, ""]
    \arrow[r, bend right=60, ""{name=D}, ""']
    & \phi'
    \arrow[Rightarrow, from=U, to=D, bend left=55, ""{name=R, below}, ""] \arrow[Rightarrow, from=U, to=D, bend right=55, ""{name=L}, ""'] \tarrow[from=L, to=R, end anchor={[yshift=0.23ex]}]{r} \arrow[phantom, from=L, to=R, end anchor={[yshift=0.6ex]}, bend left=34, ""]
\end{tikzcd}, \quad \cdots
\end{equation*}
\vspace{-0.5cm}

In this sense, higher geometry can be seen as a generalization of differential geometry which encompasses the gauge principle of physics.

In higher derived geometry, moreover, we exploit derived intersection theory to place our fields on-shell and reproduce the perturbation theory which is known to physics.
The ordinary space of field configurations is replaced by a higher (derived) geometric object which, at least infinitesimally (in a sense that can be made precise), looks like a complex of the form
\begin{equation}
   \underbrace{\cdots \xrightarrow{\;\,Q\,\;} \, E_{-3}\, \xrightarrow{\;\,Q\,\;} \, E_{-2}\,  \xrightarrow{\;\,Q\,\;} \, E_{-1}\,}_{\substack{\text{higher}\\\text{geometry}\\\text{(\textit{ghosts})}}}  \xrightarrow{\;\,Q\,\;} \!\!\!\!\underbrace{\,E_0\,}_{\substack{\text{ordinary}\\\text{geometry}\\\text{(\textit{fields})}}}\!\!\!\!  \xrightarrow{\;\,Q\,\;} \underbrace{\, E_1\,  \xrightarrow{\;\,Q\,\;} \, E_2\,  \xrightarrow{\;\,Q\,\;} \, E_3\,  \xrightarrow{\;\,Q\,\;} \cdots}_{\substack{\text{derived}\\\text{geometry}\\\text{(\textit{antifields}, \textit{antighosts})}}}
\end{equation}
where ghosts encode gauge transformations and gauge-of-gauge transformations, while antifields and antighosts are responsible for imposing the equations of motion and placing the theory on-shell (as we will see, thanks to the $L_\infty$-bracket structure).

\section{Higher geometry}

\paragraph{Simplicial sets.}
For any $n\in\mathbb{N}$, a linearly ordered set $[n] \coloneqq \{0,1,2,\dots,n\}$ with $n\in\mathbb{N}$ can be thought as an $n$\textit{-simplex}. The first ones are the following.

\vspace{-0.2cm}

{\begin{center}\includegraphics[width=11.5cm]{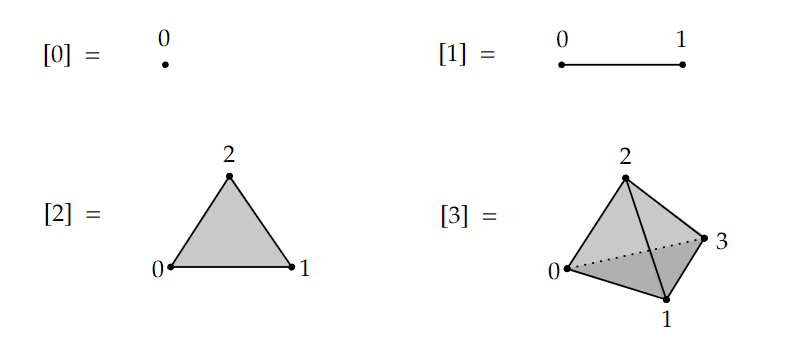}\end{center}}

\vspace{-0.2cm}

\begin{figure}[h]
    \centering
\includegraphics[width=13.5cm]{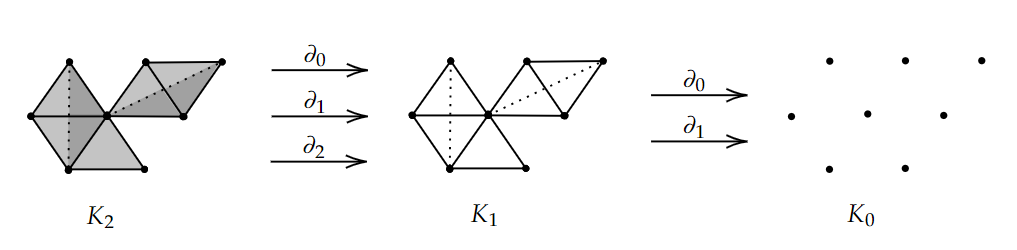}
    \caption{A simplicial set. }
    \label{fig:simplicial_set}
\end{figure}

The \textit{simplex category} $\Delta$ is the category whose objects are $n$-simplices $[n]$ for all $n\in\mathbb{N}$ and whose morphisms are order-preserving functions. In particular, we have that
\begin{itemize}
    \item  a \textit{face inclusion} $\di_i\!:[n\!-1] \hookrightarrow [n]$ for $i=0,\dots,n$ is the order-preserving injection missing $i$,
    \item  a \textit{degenerate projection} $s_i:[n+1]\twoheadrightarrow [n]$ for $i=0,\dots,n$ is the order-preserving surjection hitting $i$ twice,
\end{itemize}
Let $\mathsf{Set}$ be the category of sets, whose objects are sets and morphisms are functions.
A \textit{simplicial set} is defined as a functor
\begin{equation}
    K :\, {\Delta}^\mathrm{op}\;\longrightarrow \; \mathsf{Set}.
\end{equation}
By unravelling this definition, a simplicial set $K$ is given by a collection of sets $K_n\coloneqq K([n])$ for any $n\in\mathbb{N}$ together with the following maps:
\begin{itemize}
    \item \textit{face maps} $\partial_i\coloneqq K(\di_i):\,K_n\,\twoheadrightarrow\,K_{n-1}$ sending $n$-simplices to their $i$th boundary $(n-1)$-simplices,
    \item \textit{degeneracy maps} $\sigma_i\coloneqq K(s_i):\,K_n\,\hookrightarrow\,K_{n+1}$ sending $n$-simplices to degenerate $(n+1)$-simplices,
\end{itemize} 
which satisfy the following equations, known as \textit{simplicial identities}:
 \begin{equation}
        \begin{aligned}
            \partial_i\circ\partial_j \;&=\; \partial_{j-1}\circ\partial_i && (i<j),\\
            \partial_i\circ \sigma_j \;&=\;  \sigma_{j-1}\circ\partial_i && (i<j), \\
            \partial_i\circ  \sigma_i \;&=\; \mathrm{id} && (0\leq i \leq n), \\
            \partial_{i+1}\circ  \sigma_i \;&=\; \mathrm{id} && (0\leq i \leq n), \\
            \partial_i\circ  \sigma_j \;&=\;  \sigma_j \circ \partial_{i-1} &&(i>j+1), \\
             \sigma_i\circ \sigma_j \;&=\;  \sigma_{j+1}\circ \sigma_i &&(i\leq j).
        \end{aligned}
    \end{equation}
The collection of face maps provides a diagram (figure \ref{fig:simplicial_set}) of the following form:
\begin{equation}
   \begin{tikzcd}[row sep=scriptsize, column sep=5ex]
    \; \cdots\; \arrow[r, yshift=1.4ex] \arrow[r, yshift=2.8ex] \arrow[r] \arrow[r, yshift=-1.4ex]\arrow[r, yshift=-2.8ex] & K_3 \arrow[r, yshift=1.8ex]\arrow[r, yshift=0.6ex]\arrow[r, yshift=-1.8ex]\arrow[r, yshift=-0.6ex]& K_2
    \arrow[r, yshift=1.4ex] \arrow[r] \arrow[r, yshift=-1.4ex] & K_1  \arrow[r, yshift=0.7ex] \arrow[r, yshift=-0.7ex] & K_0 .
    \end{tikzcd}
\end{equation}

We define the \textit{category of simplicial sets} by the functor category $\mathsf{sSet} \coloneqq [\Delta^{\mathrm{op}},\mathsf{Set}]$,
whose objects are simplicial sets $K : \Delta^\mathrm{op}\rightarrow \mathsf{Set}$ and whose morphisms are natural transformations.
For any $n\in\mathbb{N}$, we define the \textit{$n$-simplex simplicial set} $\Delta^n\in\mathsf{sSet}$ as the Yoneda embedding of the $n$-simplex $[n]$ into the category of simplicial sets $\Delta^n \coloneqq \mathrm{Hom}_{\Delta}(-,[n])$.
The category $\sSet$ of simplicial sets is naturally a simplicial category, i.e. a category enriched over $\sSet$ itself. In fact, the \textit{internal hom} $[X,Y]$ of two simplicial sets $X$ and $Y$ is a simplicial set given by
\begin{equation}
    [X,Y]\,: [n] \;\longrightarrow\; \Hom_\mathsf{sSet}(X\times \Delta^n, \,Y),
\end{equation}
where $\Delta^n\in\mathsf{sSet}$ is the $n$-simplex simplicial set.

\paragraph{$\infty$-groupoids.}
A \textit{Kan complex} is a simplicial set $K\in\mathsf{sSet}$ that satisfies the \textit{Kan condition}: that is, for any image of a horn $f:\Lambda^n_i\rightarrow K$ with $0\leq i\leq n$ in $K$, the missing $i$th face must be in $K$ too, i.e. there must exist a map $f'$ such that
\begin{equation}\label{eq:kan}
    \begin{tikzcd}[row sep=5.5ex, column sep=6.5ex]
    \Lambda^n_i \arrow[r, "f"]\arrow[d, hook]& K \,. \\
    \Delta^n \arrow[ur, "f'"']
    \end{tikzcd}  
\end{equation}
A \textit{weak Kan complex} is a simplicial set $K\in\mathsf{sSet}$ that satisfies the Kan condition only for the internal $n$-horns $\Lambda_i^n$, i.e. those with $0<i<n$. 
In this article we will adopt the following model: an \textit{$\infty$-groupoid} will be a Kan complex and an \textit{$(\infty,1)$-category} will be a weak Kan complex.

Roughly speaking, a \textit{model category} is a category with three distinguished classes of morphisms called weak equivalences, fibrations and cofibrations satisfying certain axioms which relate them.
We will denote by $\sSet_{\mathsf{Quillen}}$ the simplicial category of simplicial sets equipped with Quillen model structure \cite{Quillen:1967ha}, whose weak equivalences are weak homotopy equivalences of simplicial sets and whose fibrations are Kan fibrations. 
It can be shown that the full subcategory $\sSet_{\mathsf{Quillen}}^\circ$ of fibrant-cofibrant objects of $\sSet_{\mathsf{Quillen}}$ is equivalent to the simplicial-category of Kan complexes, i.e.
\begin{equation}
    \mathsf{KanCplx} \;\simeq\; \sSet_{\mathsf{Quillen}}^\circ.
\end{equation}
Now, we can make this simplicial category into a fully fledged $(\infty,1)$-category. Essentially, an $(\infty,1)$-category is a simplicial set which satisfies the weak Kan condition.
It is a standard technique \cite[Section 1.1.5]{topos} that, by applying the well-known homotopy-coherent nerve functor $\mathrm{N}_{hc}$ to our simplicial category, one obtains the $(\infty,1)$-category of $\infty$-groupoids, i.e.
\begin{equation}
    \mathbf{\infty Grpd} \;\coloneqq\; \mathrm{N}_{hc}(\sSet_{\mathsf{Quillen}}^\circ).
\end{equation}

\begin{figure}[h]
    \centering
\includegraphics[width=12cm]{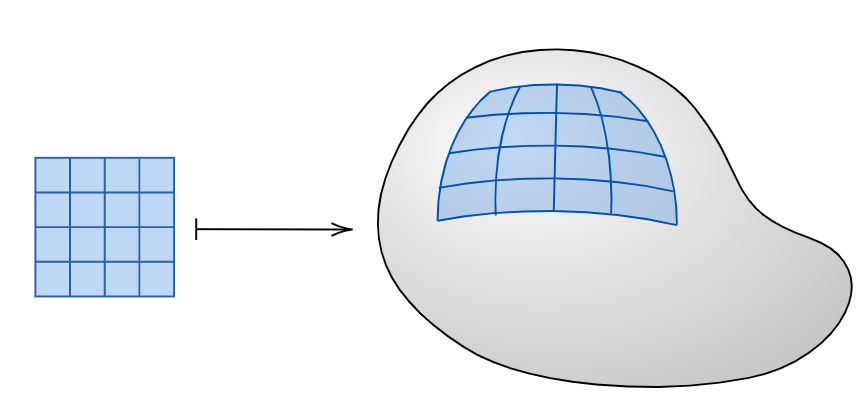}
    \caption{Smooth stacks, visualized as geometric objects probed by ordinary manifolds.}
    \label{fig:my_label1}
\end{figure}

There exists an embedding $\mathbf{B}$, which we call \textit{delooping}, which identifies a smooth $\infty$-group $G$ with a smooth $\infty$-groupoid $\mathbf{B}G\in \mathbf{H}$ whose set of $0$-simplices is a singleton, i.e.
\begin{equation}
    \mathbf{B}G\;=\; \left( \begin{tikzcd}[row sep=scriptsize, column sep=5ex]
    \; \cdots\; \arrow[r, yshift=1.4ex] \arrow[r, yshift=2.8ex] \arrow[r] \arrow[r, yshift=-1.4ex]\arrow[r, yshift=-2.8ex] & G_3 \arrow[r, yshift=1.8ex]\arrow[r, yshift=0.6ex]\arrow[r, yshift=-1.8ex]\arrow[r, yshift=-0.6ex]& G_2
    \arrow[r, yshift=1.4ex] \arrow[r] \arrow[r, yshift=-1.4ex] & G_1  \arrow[r, yshift=0.7ex] \arrow[r, yshift=-0.7ex] & \ast 
    \end{tikzcd}    \right).
\end{equation}
For example, if $G$ is an ordinary group, the corresponding simplicial set is given by $[n]\mapsto G^{\times n}$. More precisely, this is given by the following diagram:
\begin{equation*}
    \mathbf{B}G\;=\; \left( \begin{tikzcd}[row sep=scriptsize, column sep=14ex]
    \, \cdots\;\; G\times G\times G \arrow[r, yshift=3.8ex, "{(\pi_1,\pi_2)}"]\arrow[r, yshift=1.4ex, "{((-)\cdot(-),\,\pi_3)}"]\arrow[r, yshift=-1.4ex, "{(\pi_1,\,(-)\cdot(-))}"]\arrow[r, yshift=-3.8ex, "{(\pi_2,\pi_3)}"] & G\times G
    \arrow[r, yshift=3.2ex, "\pi_1"] \arrow[r, "(-)\cdot(-)"] \arrow[r, yshift=-3.2ex, "\pi_2"] & G  \arrow[r, yshift=1.4ex, "\ast"] \arrow[r, yshift=-1.4ex, "\ast"] & \ast 
    \end{tikzcd}    \right),
\end{equation*}
where $\pi_i:G^{\times n}\rightarrow G$ is the projection of the $i$th copy and $(-)\cdot(-)$ is the group product of $G$.

\paragraph{Smooth stacks.} 
If a category $\mathsf{C}$ has the structure of a \textit{site with enough points}, there exists a model structure $[\mathsf{C},\sSet]_{\mathsf{proj,loc}}$ which is known as the \textit{projective local model structure} \cite{Blander2001LocalPM} and whose local weak equivalences are natural transformations which are stalk-wise weak homotopy equivalences of simplicial sets. Since the category $\Mfd$ of smooth manifolds has enough points, there is a simplicial model category $[\mathsf{Mfd}^\op,\mathsf{sSet}]_{\mathsf{proj,loc}}$.
By using the homotopy coherent nerve functor $\mathbf{N}_{hc}$, which maps simplicial categories to simplicial sets, it is possible to convert our simplicial model category into an $(\infty,1)$-category
\begin{equation}
    \topos \;=\; \mathbf{N}_{hc}([\mathsf{Mfd}^\op,\mathsf{sSet}]_{\mathsf{proj,loc}}^\circ).
\end{equation}
Given a covering family $\{U_i\rightarrow U\}_{i\in I}$, it is possible to construct a simplicial object known as \v{C}ech nerve of the smooth manifold $U$ by
\begin{equation}
    \check{C}(U)_\bullet \;=\;  \bigg( \begin{tikzcd}[row sep=scriptsize, column sep=3.5ex]
    \; \cdots\; \arrow[r, yshift=1.8ex]\arrow[r, yshift=0.6ex]\arrow[r, yshift=-1.8ex]\arrow[r, yshift=-0.6ex]& \displaystyle\coprod_{i,j,k\in I}U_i\times_U U_j\times_U U_k 
    \arrow[r, yshift=1.4ex] \arrow[r] \arrow[r, yshift=-1.4ex] & \displaystyle\coprod_{i,j\in I}U_i\times_U U_j  \arrow[r, yshift=0.7ex] \arrow[r, yshift=-0.7ex] & \displaystyle\coprod_{i\in I}U_i 
    \end{tikzcd}    \bigg),
\end{equation}
whose colimit is the original smooth manifold $U \simeq \mathrm{co}\!\lim_{[n]\in\Delta}\check{C}(U)_n$.
By unravelling the definition of a smooth stack, more concretely, one has that it is a functor $X:\Mfd \longrightarrow \sSet$, i.e. so that
\begin{equation*}
    U\,\mapsto\, X(U)\,=\, \left( \begin{tikzcd}[row sep=scriptsize, column sep=5ex]
    \; \cdots\; \arrow[r, yshift=1.4ex] \arrow[r, yshift=2.8ex] \arrow[r] \arrow[r, yshift=-1.4ex]\arrow[r, yshift=-2.8ex] & X_3(U) \arrow[r, yshift=1.8ex]\arrow[r, yshift=0.6ex]\arrow[r, yshift=-1.8ex]\arrow[r, yshift=-0.6ex]& X_2(U)
    \arrow[r, yshift=1.4ex] \arrow[r] \arrow[r, yshift=-1.4ex] & X_1(U)  \arrow[r, yshift=0.7ex] \arrow[r, yshift=-0.7ex] & X_0(U) 
    \end{tikzcd}    \right),
\end{equation*} 
satisfying the following properties:
\begin{enumerate}
    \item \textit{object-wise fibrancy}: for any $U\in\mathsf{Mfd}$, the simplicial set $X(U)$ is Kan-fibrant;
    \item \textit{pre-stack condition}: for any diffeomorphism $U\xrightarrow{\,\simeq\,} U'$, the induced morphism $X(U')\rightarrow X(U)$ is an equivalence of simplicial sets;
    \item \textit{descent condition}: for any \v{C}ech nerve $\check{C}(U)_\bullet\rightarrow U$, the natural morphism
    \begin{equation}
    X(U) \;\longrightarrow\; \lim_{[n]\in\Delta} \bigg(\prod_{i_1,\dots,i_n\in I} \! X(U_{i_1\!}\times_{U\!}\cdots_{\!}\times_{U\!}U_{i_n})\,\bigg)
    \end{equation}
    is an equivalence of simplicial sets.
\end{enumerate}

\section{Higher gauge theories}

Higher gauge theories \cite{baez2005higher} are a generalization of ordinary gauge theories, where the usual connection $1$-forms are extended to higher differential forms. This is obtained by generalizing ordinary principal bundles to principal $\infty$-bundles, where the fiber is a general smooth $\infty$-group.
One way to think about higher gauge theories is as a higher-dimensional generalization of ordinary gauge theories. In fact, in ordinary gauge theory, the physical objects of interest are (electrically) charged point-particles, and their coupling to the gauge field is described by the parallel transport of the connection (a local $1$-form) on the corresponding principal bundle. In higher gauge theory, the physical objects of interest are charged strings or branes, and the coupling of these objects to the higher gauge field is described by the parallel transport of a higher connection (a local higher-form) on the corresponding higher principal bundles.

\subsection{Principal $\infty$-bundles with connections}

Early work on categorified bundles include \cite{bartels2006higher, jardine_luo_2006, 2010arXiv1009.2930W, Wockel_2011}.
The theory of principal $\infty$-bundles in the context of $(\infty,1)$-categories was developed by \cite{Principal1,Principal2}. 
For an overview on principal $\infty$-bundles, we redirect to the article \cite{bunk2023inftybundles}.

\paragraph{Principal $\infty$-bundles.}
Let us start from ordinary differential geometry. In this context, for a fixed manifold $M$ and ordinary Lie group $G$, a principal $G$-bundle on $M$ is topologically given by an element of the $1$st non-abelian $G$-cohomology group $\mathrm{H}^1(M,G)$. In fact, this will be an equivalence class $[g_{\alpha\beta}]$ represented by a \v{C}ech $G$-cocycle $g_{\alpha\beta}\in\Coo(U_\alpha\cap U_\beta,G)$, where the equivalence relation is given by \v{C}ech coboundaries $g_{\alpha\beta}\sim\eta_{\alpha} g_{\alpha\beta}\eta_{\beta}^{-1}$ for $\eta_{\alpha}\in\Coo(U_\alpha,G)$. 
We want to refine this description to the context of smooth stacks.
The groupoid $\mathbf{H}(M,\mathbf{B}G)$ has, as $0$-simplices, all nonabelian \v{C}ech $G$-cocycles $g_{\alpha\beta}$ on $M$ and, as $1$-simplices, all couboundaries $g_{\alpha\beta}\mapsto\eta_{\alpha} g_{\alpha\beta}\eta_\beta^{-1}$ between them. 
Schematically, we have:
\begin{equation}
    \mathbf{H}(M,\mathbf{B}G)\,\simeq\,\left\{\begin{tikzcd}[row sep=scriptsize, column sep=12ex]
    M \arrow[r, bend left=50, ""{name=U, below}, "g_{\alpha\beta}"]
    \arrow[r, bend right=50, "\eta_{\alpha} g_{\alpha\beta}\eta_{\beta}^{-1}"', ""{name=D}]
    & \mathbf{B}G
    \arrow[Rightarrow, from=U, to=D, "\eta_{\alpha}"]
\end{tikzcd} \right\}.
\end{equation}
In geometric terms, the $0$-simplices are all the principal $G$-bundles on $M$ and the $1$-simplices are all the isomorphisms (i.e. gauge transformations) between them. Thus, we can see a principal $G$-bundle as an $0$-simplex of the $\infty$-groupoid $\mathbf{H}(M,\mathbf{B}G)$.
To recover the ordinary picture, we only need to take the set of connected components of the $\infty$-groupoid of ordinary principal bundles:
\begin{equation}\label{eq:pathconnected}
    \mathrm{H}^1(M,G) \;=\; \pi_0\mathbf{H}(M,\mathbf{B}G).
\end{equation}
To any cocycle $g:M \rightarrow \mathbf{B}G$ is associated a principal $G$-bundle $\pi:P\twoheadrightarrow M$ given by pullback
\begin{equation}\label{eq:firstbundle}
    \begin{tikzcd}[row sep=7.5ex, column sep=7.5ex]
    P \arrow[d, "\mathrm{hofib}(g)"']\arrow[r] & \ast \arrow[d] \\
    M \arrow[r, "g"] & \mathbf{B}G ,
\end{tikzcd}
\end{equation}
where we called $\pi=:\mathrm{hofib}(g)$ the projection of the bundle.
The fundamental idea for defining principal $\infty$-bundles is to let the formalism above work for general group objects in $\topos$.

To achieve this, we must first define what is meant by a \textit{$G$-action} on a stack $P\in\topos$ for a given smooth $\infty$-group $G\in\topos$. This is nothing but a simplicial diagram of stacks
\begin{equation}
    \begin{tikzcd}[row sep=scriptsize, column sep=5ex]
    \; \cdots\;  \arrow[r, yshift=1.8ex]\arrow[r, yshift=0.6ex]\arrow[r, yshift=-1.8ex]\arrow[r, yshift=-0.6ex]& P\times G\times G
    \arrow[r, yshift=1.4ex] \arrow[r] \arrow[r, yshift=-1.4ex] & P\times G  \arrow[r, yshift=0.7ex, "\partial_0"] \arrow[r, yshift=-0.7ex, "\partial_1"'] & P,
    \end{tikzcd} 
\end{equation}
such that $\partial_1$ is the natural projection and the degree-wise natural projections
\begin{equation}
    \begin{tikzcd}[row sep=6.5ex, column sep=5ex]
    \; \cdots\;  \arrow[r, yshift=1.8ex]\arrow[r, yshift=0.6ex]\arrow[r, yshift=-1.8ex]\arrow[r, yshift=-0.6ex]& P\times G\times G
    \arrow[d]\arrow[r, yshift=1.4ex] \arrow[r] \arrow[r, yshift=-1.4ex] & P\times G  \arrow[d]\arrow[r, yshift=0.7ex] \arrow[r, yshift=-0.7ex] & P\arrow[d]\\
        \; \cdots\;  \arrow[r, yshift=1.8ex]\arrow[r, yshift=0.6ex]\arrow[r, yshift=-1.8ex]\arrow[r, yshift=-0.6ex]& G\times G
    \arrow[r, yshift=1.4ex] \arrow[r] \arrow[r, yshift=-1.4ex] & G  \arrow[r, yshift=0.7ex] \arrow[r, yshift=-0.7ex] & \ast 
    \end{tikzcd}
\end{equation}
give a morphism of simplicial objects.
Now, a \textit{principal $\infty$-bundle} $P\rightarrow M$ can be defined as the datum of
\begin{enumerate}
    \item[(\textit{i})] a morphism $P\rightarrow M$ in $\topos$,
    \item[(\textit{ii})] a $G$-action on $P$.
\end{enumerate}
which are compatible in the sense that the morphism $P\rightarrow M\simeq P/\!/G$ is the quotient map of the action.
Crucially, in \cite{Principal1}, it is proved that, for any map $g:M\rightarrow \B G$ with $G$ a smooth $\infty$-groupoid, its homotopy fiber $\mathrm{hofib}(g):P\rightarrow M$ is canonically a $G$-principal $\infty$-bundle.
This means that any principal $\infty$-bundle is given by a pullback diagram of stacks
\begin{equation}
    \begin{tikzcd}[row sep=7.5ex, column sep=7.5ex]
    P \arrow[d, "\mathrm{hofib}(g)"']\arrow[r] & \ast \arrow[d] \\
    M \arrow[r, "g"] & \mathbf{B}G
\end{tikzcd}
\end{equation}
for some cocycle $g\in\mathbf{H}(M,\mathbf{B}G)$ and vice versa.
Thus, just like in the ordinary case, we can simply define the $\infty$-groupoid of principal $G$-bundles on a $M$ by
\begin{equation}\label{eq:groupoidofprinc}
    \mathrm{Bun}_G(M) \;\coloneqq \; \mathbf{H}(M,\mathbf{B}G).
\end{equation}
For this reason, $\mathbf{B}G$ is also known as \textit{moduli stack of principal $G$-bundles}.


\paragraph{Connections on $\infty$-bundles.}
Now, to formalize gauge theories, we want to introduce the moduli stack of principal bundle with connection $\mathbf{B}G_{\mathrm{conn}}$, which refines the moduli stack of principal bundles $\mathbf{B}G$.
In other words, we want a diagram of the form
\begin{equation}
    \begin{tikzcd}[row sep={11ex,between origins}, column sep={6ex}]
    & \mathbf{B}G_{\mathrm{conn}} \arrow[d, "\mathrm{forget}"] \\
    M \arrow[r, "g"]\arrow[ur, "{(A,g)}"] & \mathbf{B}G,
    \end{tikzcd}
\end{equation}
where a map $(A,g):M\rightarrow \mathbf{B}G_{\mathrm{conn}}$ would encode both the geometric data of a principal bundle $g:M\rightarrow \mathbf{B}G$ and the differential data of a principal connection $A$ on such a bundle.

Let us first see what happens for ordinary principal bundles, where $G$ is an ordinary Lie group. 
In this case, a map $(A_{\alpha },\,g_{\alpha\beta }):M \rightarrow \mathbf{B}G_\mathrm{conn}$ provides the \v{C}ech cocycle encoding the transition functions $(g_{\alpha\beta})$ of a principal $G$-bundle with a connection $(A_\alpha)$, i.e. local $1$-forms $A_\alpha\in\Omega^1(U_\alpha,\mathfrak{g})$ glued by $A_\beta = g_{\alpha\beta }^{-1}(A_{\alpha }+\di)g_{\alpha\beta }$.
Morphisms between such maps are nothing but gauge transformations
\begin{equation}
    (A_{\alpha },\,g_{\alpha\beta }) \;\xmapsto{\quad \eta_{\alpha } \quad}\; (A'_{\alpha },\,g'_{\alpha\beta })=\big(\eta_{\alpha }^{-1}(A_{\alpha }+\di)\eta_{\alpha },\; \eta_{\alpha }g_{\alpha\beta }\eta_{\beta }^{-1}\big) ,
\end{equation}
where the gauge parameter $\eta_{\alpha }\in\Coo(U_\alpha,G)$ is a local $G$-valued function.

The problem of systematizing the construction above for general smooth $\infty$-groups was discussed in \cite{schreiber2013connections, 2017arXiv170408542W}.
Such a construction has to generalize ordinary parallel transport to principal $\infty$-bundles.
First, one can define the \textit{path $\infty$-groupoid} $\mathscr{P}(M)$, a smooth stack whose $0$-simplices are points of $M$, $1$-simplices paths between points of $M$, $2$-simplices smooth deformations of paths, and so on.
By parallel transport, it is possible to show that a map $\mathscr{P}(M)\rightarrow \B G$ is equivalently a principal $G$-bundle on $M$ equipped with a flat connection.
To obtain a general connection, one needs to introduce the \textit{$\infty$-group of inner automorphisms}\footnote{This is also known as groupal model for universal principal $\infty$-bundles.} by the short exact sequence
\begin{equation}
    \,G\, \longrightarrow \,\mathrm{Inn}(G)\, \longrightarrow \,\B G .
\end{equation} 
However, a map $\mathscr{P}(U) \rightarrow \mathbf{B}\mathrm{Inn}(G)$ still does not provide general connections (it does only for ordinary Lie groups and abelian $\infty$-groups). For a general smooth $\infty$-group $G$, such a map would produce vanishing higher-form curvatures (\textit{fake-flatness}). This issue was observed by \cite{baez2005higher} and a solution was proposed by \cite{Saem17, Kim:2019owc, Borsten:2021ljb, Rist:2022hci}. 
The smooth stack $\mathbf{B}G_{\mathrm{conn}}$ must be constructed by using an adjusted notion of parallel transport, which makes use of an adjusted version $\mathrm{Inn}_{\mathrm{adj}}(G)$ of the $\infty$-group of inner automorphisms.
So, given a smooth $\infty$-group $G$, the \textit{moduli stack $\mathbf{B}G_\mathrm{conn}$ of $G$-bundles with connection} can be constructed from the functor
\begin{equation}
    \begin{aligned}
        \mathbf{B}G_{\mathrm{conn}}:\, \mathsf{Mfd}^\mathrm{op} \; &\longrightarrow \; \mathbf{\infty Grpd} \\
        U \; &\longmapsto \; \mathbf{H}\big(\mathscr{P}(U),\,\mathbf{B}\mathrm{Inn}_{\mathrm{adj}}(G)\big),
    \end{aligned}
\end{equation}
where $\mathscr{P}(U)$ is the path $\infty$-groupoid of the smooth manifold $U$ and $\mathrm{Inn}_{\mathrm{adj}}(G)$ is the \textit{adjusted $\infty$-group of inner automorphisms} of $G$. For more details and discussion, see \cite{Kim:2019owc, Rist:2022hci}.

\paragraph{BRST $L_\infty$-algebras.}
Now, we will discuss how principal bundles on a manifold $M$ with connection, which are equivalently maps $\mathscr{P}(M)\rightarrow \mathbf{B}\mathrm{Inn}_\mathrm{adj}(G)$, give rise to BRST $L_\infty$-algebras. 
Notice that the infinitesimal version of such a map is of the form
\begin{equation}
    A:\; TM \;\longrightarrow\; \B\mathfrak{inn}_\mathrm{adj}(\mathfrak{g}).
\end{equation}
Given the identification $\CE(TM)=(\Omega^\bullet(M),\di)$ and the definition $\mathrm{W}_\mathrm{adj}(\mathfrak{g})\coloneqq \CE(\mathfrak{inn}_\mathrm{adj}(\mathfrak{g}))$, this is equivalently a map of the corresponding Chevalley-Eilenberg dg-algebras
\begin{equation}
    (\Omega^\bullet(M),\di) \;\longleftarrow\; \mathrm{W}_\mathrm{adj}(\mathfrak{g}) \;:A.
\end{equation}
Such a map is nothing but a $\mathfrak{g}$-valued differential form $A$ with non-necessarily-vanishing curvature 
\begin{equation}
    F \;=\; \di A + \ell_1(A) + \frac{1}{2}\ell_2(A\,\overset{\wedge}{,}\,A)+ \frac{1}{3!}\ell_3(A\,\overset{\wedge}{,}\,A\,\overset{\wedge}{,}\,A) + \dots
\end{equation}
The cochain complex of such maps can be constructed by $E \coloneqq \mathrm{Hom}_{\leq 0}(\mathrm{W}_\mathrm{adj}(\mathfrak{g}),\,\Omega^\bullet(M))$, which is
\begin{equation}
   \underbrace{\cdots \xrightarrow{\;\,\di+\ell_1\,\;} \, E_{-3}\, \xrightarrow{\;\,\di+\ell_1\,\;} \, E_{-2}\,  \xrightarrow{\;\,\di+\ell_1\,\;} \, E_{-1}\,}_{\substack{\text{higher}\\\text{geometry}\\\text{(\textit{ghosts})}}}  \xrightarrow{\;\,\di+\ell_1\,\;} \!\!\!\!\underbrace{\,E_0\,}_{\substack{\text{ordinary}\\\text{geometry}\\\text{(\textit{fields})}}}\!\!\!\!  \xrightarrow{\;\,\;0\;\,\;} 0
\end{equation}
where $E_{-i} \cong \bigoplus_{n=0}^{\mathrm{dim}(M)\!} \Omega^n(M)\otimes \mathfrak{g}_{-i-n}$ for any $i\in\mathbb{N}$. The shifted complex $\mathfrak{L}\coloneqq E[-1]$ can be equipped with a natural $L_\infty$-algebra structure, whose $k$-bracket is given by $\ell_k(-\,\overset{\wedge}{,}\,-\,\overset{\wedge}{,}\,\cdots\,\overset{\wedge}{,}\,-)$, where $\ell_k$ is the $k$-bracket of $\mathfrak{g}$.
Elements $A\in E_0$ are (topologically trivial) higher gauge fields, while elements $\eta\in E_{-1}$ are ghost fields which parameterize infinitesimal gauge transformations by
\begin{equation}
    \delta_\eta A \;=\; \di\eta + \ell_1(\eta) + \frac{1}{2}\ell_2(\eta\,\overset{\wedge}{,}\,A) + \frac{1}{3!}\ell_3(\eta\,\overset{\wedge}{,}\, A\,\overset{\wedge}{,}\,A) + \dots
\end{equation}

\paragraph{Example: Yang-Mills theory}
Consider the (local) $L_\infty$-algebra $\mathfrak{L}$, whose underlying complex is the differential graded vector space 
\begin{equation}\label{eq:brst}
\begin{aligned}
    \mathfrak{L}[1] \;=\; \,&\Big( \begin{tikzcd}[row sep={14.5ex,between origins}, column sep= 5ex]
    \Omega^0(M,\mathfrak{g}) \arrow[r, "\di"] & \Omega^1(M,\mathfrak{g})
    \end{tikzcd}  \Big)\\
    {\scriptstyle\text{deg}\,=} &\quad \;\;\begin{tikzcd}[row sep={14.5ex,between origins}, column sep= 5ex]
    {\scriptstyle -1} && \quad \;\;{\scriptstyle 0}
    \end{tikzcd} ,
\end{aligned}
\end{equation}
and whose $L_\infty$-bracket structure has only the following non-trivial brackets:
\begin{equation}
    \begin{aligned}
        \ell_1(c) \;&=\; \di c, \\
        \ell_2(c_1,c_2) \;&=\; [c_1,c_2]_\mathfrak{g}, \\
        \ell_2(c,A) \;&=\; [c,A]_\mathfrak{g},
    \end{aligned}
\end{equation}
for any elements $c_k\in\Omega^0(M,\mathfrak{g})$ and $A\in\Omega^1(M,\mathfrak{g})$.
Thus, the $L_\infty$-algebra $\mathfrak{L}$ is precisely the algebraic incarnation of the BRST complex of physics.

\subsection{Bundle gerbes and type II string theory}

Abelian bundle gerbes are a categorification of principal $\mathrm{U}(1)$-bundles introduced by \cite{Murray, Murray2}. More recently, in \cite{Principal1}, bundle gerbes have been formalized as a special case of principal $\infty$-bundle, where the structure smooth $\infty$-group is $G=\mathbf{B}\mathrm{U}(1)$, the so-called circle $2$-group. 
For a review, see \cite{Bunk:2021quu}. 
We will see that the connection of a bundle gerbe, as a higher gauge field, is nothing but the Kalb-Ramond field of type II string theory.

\paragraph{Bundle gerbe as principal $\infty$-bundle.}
More generally, a \textit{bundle $n$-gerbe} $P\twoheadrightarrow M$ is defined as a principal $\mathbf{B}^n\mathrm{U}(1)$-bundle on some smooth manifold $M$, for any $n\in\mathbb{N}$. 
Thus, the $\infty$-groupoid of bundle $n$-gerbes on a fixed smooth manifold $M$ is
\begin{equation}
    n\text{-}\mathrm{Gerb}(M) \;\coloneqq \; \mathbf{H}(M,\mathbf{B}^{n+1}\mathrm{U}(1)).
\end{equation}
We can notice that bundle $n$-gerbes on a base manifold $M$ are topologically classified by the $n$th cohomology group of $M$, in fact by taking the group of path-connected components we have
\begin{equation}
    \pi_0\mathbf{H}\big(M,\mathbf{B}^{n+1}\mathrm{U}(1)\big)  \;\cong\; \mathrm{H}^{n+1}(M,\mathrm{U}(1))  \;\cong\;  \mathrm{H}^{n+2}(M,\mathbb{Z}).
\end{equation}
We call a bundle $1$-gerbe simply \textit{bundle gerbe}.
Therefore, bundle gerbes $P\rightarrow M$ on a base manifold $M$ are topologically classified by an element $\mathrm{dd}(P)\in \mathrm{H}^3(M,\mathbb{Z})$, which is called their \textit{Dixmier-Douady class}. 
This generalizes the $1$st Chern class in $\mathrm{H}^2(M,\mathbb{Z})$ classifying ordinary circle bundles.

\paragraph{Bundle gerbe in \v{C}ech data.}
Let $\mathcal{U}\coloneqq \{U_\alpha\}$ be a good cover for the base manifold $M$. The \v{C}ech groupoid $\check{C}(\mathcal{U})$ is defined as the smooth stack represented by the following simplicial manifold
\begin{equation}\label{eq:cechgroupoidsimplicial}
    \begin{tikzcd}[row sep=scriptsize, column sep=6ex] \cdots\arrow[r, yshift=1.8ex]\arrow[r, yshift=0.6ex]\arrow[r, yshift=-0.6ex]\arrow[r, yshift=-1.8ex] & \bigsqcup_{\alpha\beta\gamma}U_{\alpha}\cap U_\beta\cap U_\gamma\arrow[r, yshift=1.4ex]\arrow[r]\arrow[r, yshift=-1.4ex]& \bigsqcup_{\alpha\beta}U_{\alpha}\cap U_\beta  \arrow[r, yshift=0.7ex] \arrow[r, yshift=-0.7ex] & \; \bigsqcup_{\alpha} U_{\alpha} .
    \end{tikzcd}
\end{equation}
The pre-sheaf represented by a smooth manifold $M$ is not generally a cofibrant object in the model category of smooth stacks. However, there is a natural weak equivalence between the \v{C}ech groupoid $\check{C}(\mathcal{U})$, which is cofibrant, and the pre-sheaf represented by $M$, so that we work with the former.
Thus, we can write the map between $M$ and the moduli stack $\mathbf{B}^2\mathrm{U}(1)$ as a morphism of simplicial manifolds $f: \check{C}(\mathcal{U}) \rightarrow \mathbf{B}^2\mathrm{U}(1)$. By using the definition of the \v{C}ech groupoid and $\mathbf{B}^2\mathrm{U}(1)$, we obtain the following diagram:
\begin{equation}
    \begin{tikzcd}[row sep={10ex}, column sep={3ex}]
     \bigsqcup_{\alpha\beta\gamma\delta}U_{\alpha}\cap U_\beta\cap U_\gamma\cap U_\delta \arrow[d]\arrow[r, yshift=1.8ex]\arrow[r, yshift=0.6ex]\arrow[r, yshift=-0.6ex]\arrow[r, yshift=-1.8ex] &\bigsqcup_{\alpha\beta\gamma}U_{\alpha}\cap U_\beta\cap U_\gamma\arrow[d] \arrow[r, yshift=1.4ex]\arrow[r]\arrow[r, yshift=-1.4ex] & \bigsqcup_{\alpha\beta}U_{\alpha}\cap U_\beta  \arrow[d]\arrow[r, yshift=0.7ex] \arrow[r, yshift=-0.7ex] & \; \bigsqcup_{\alpha} U_{\alpha} \arrow[d] \\
    \mathrm{U}(1)\times\mathrm{U}(1)\arrow[r, yshift=1.8ex]\arrow[r, yshift=0.6ex]\arrow[r, yshift=-0.6ex]\arrow[r, yshift=-1.8ex] & \mathrm{U}(1)\arrow[r, yshift=1.4ex]\arrow[r]\arrow[r, yshift=-1.4ex] & \;\ast\;  \arrow[r, yshift=0.7ex] \arrow[r, yshift=-0.7ex] & \; \ast 
    \end{tikzcd}
\end{equation}
More in detail, a morphism is given by a collection $(G_{\alpha\beta\gamma})$ of local scalars on threefold overlaps of patches $U_\alpha\cap U_\beta\cap U_\gamma$ satisfying the cocycle condition
\begin{equation}
    \begin{aligned}
        G_{\alpha\beta\gamma}-G_{\beta\gamma\delta}+G_{\gamma\delta\alpha}-G_{\delta\alpha\beta}\;\in\; 2\pi\mathbb{Z},
    \end{aligned}
\end{equation}
on fourfold overlaps of patches. The $1$-morphisms between these objects are \v{C}ech coboundaries (i.e. the gauge transformations of the bundle gerbe) given by collections $(\eta_{\alpha\beta })$ of local scalars on overlaps $U_\alpha\cap U_\beta$ so that
\begin{equation}
    \begin{aligned}
        G_{\alpha\beta\gamma} \;\mapsto\; G_{\alpha\beta\gamma} + \eta_{\alpha\beta }+\eta_{\beta\gamma }+\eta_{\gamma\alpha }
    \end{aligned}
\end{equation}
The $2$-morphisms between $1$-morphisms (i.e. the gauge-of-gauge transformations of the bundle gerbe) are given by collections $(\epsilon_{\alpha })$ of local scalars on each $U_\alpha$ so that
\begin{equation}
    \begin{aligned}
        \eta_{\alpha\beta } \;\Mapsto\; \eta_{\alpha\beta }+\epsilon_{\alpha }-\epsilon_{\beta } .
    \end{aligned}
\end{equation}
In terms of diagrams, we can write this $2$-groupoid of abelian bundle gerbes as follows:
\begin{equation}
    \mathbf{H}\big(M,\mathbf{B}^2\mathrm{U}(1)\big) \,\simeq\, \left\{\; \begin{tikzcd}[row sep=scriptsize, column sep=26ex]
    \;\;\; M \arrow[r, bend left=60, ""{name=U, below}, "(G_{(\alpha\beta\gamma)})"]
    \arrow[r, bend right=60, "(G'_{(\alpha\beta\gamma)})"', ""{name=D}]
    & \qquad\,\mathbf{B}^2\mathrm{U}(1)
    \arrow[Rightarrow, from=U, to=D, bend left=55, "(\eta'_{\alpha\beta })", ""{name=R, below}] \arrow[Rightarrow, from=U, to=D, bend right=55, "(\eta_{\alpha\beta })"', ""{name=L}] \tarrow[from=L, to=R, end anchor={[yshift=0.6ex]}]{r} \arrow["(\epsilon_{\alpha })", phantom, from=L, to=R, end anchor={[yshift=0.6ex]}, bend left=34]
\end{tikzcd}\right\}
\end{equation}

\paragraph{B-field as connection on a bundle gerbe.} Let us now construct a bundle gerbe equipped with connection, which globally formalizes a Kalb-Ramond field. For abelian bundle gerbes, the construction of the stack $\mathbf{B}^2 \mathrm{U}(1)_{\mathrm{conn}}$ is particularly simple. An abelian {bundle gerbe with connection} is given by a map $M\rightarrow\mathbf{B}^2\mathrm{U}(1)_{\mathrm{conn}}$. Let $\mathcal{U}=\{U_\alpha\}$ be a good cover of $M$.
An object of the $\infty$-groupoid $\mathbf{H}(M,\mathbf{B}^2\mathrm{U}(1)_{\mathrm{conn}})$ is given, in \v{C}ech-Deligne data, by a collection $(B_{\alpha },\Lambda_{\alpha\beta },G_{\alpha\beta\gamma})$ of $2$-forms $B_{\alpha }\in\Omega^2(U_\alpha)$, $1$-forms $\Lambda_{\alpha\beta }\in\Omega^2(U_\alpha\cap U_\beta)$ and scalars $G_{\alpha\beta\gamma}\in\Coo(U_\alpha\cap U_\beta\cap U_\gamma)$ respectively on patches, two-fold overlaps and three-fold overlaps, such that they satisfy the gluing conditions
\begin{equation}
    \begin{aligned}
        B_{\beta }-B_{\alpha } &= \mathrm{d}\Lambda_{\alpha\beta }, \\
        \Lambda_{\alpha\beta }+\Lambda_{\beta\gamma }+\Lambda_{\gamma\alpha } &= \mathrm{d}G_{\alpha\beta\gamma} \\
        G_{\alpha\beta\gamma}-G_{\beta\gamma\delta}+G_{\gamma\delta\alpha}-G_{\delta\alpha\beta}&\in 2\pi\mathbb{Z}.
    \end{aligned}
\end{equation}
The $1$-morphisms between these objects are \v{C}ech-Deligne coboundaries (in physical words the gauge transformations of the bundle gerbe), given by collections $(\eta_{\alpha},\eta_{\alpha\beta })$ of local $1$-forms $\eta_{\alpha }\in\Omega^1(U_\alpha)$ and local scalars $\eta_{\alpha\beta }\in\Coo(U_\alpha\cap U_\beta)$, so that
\begin{equation}\label{eq:coboundaries}
    \begin{aligned}
        B_{\alpha } &\mapsto B_{\alpha } + \mathrm{d}\eta_{\alpha}, \\
        \Lambda_{\alpha\beta } &\mapsto \Lambda_{\alpha\beta }+\eta_{\alpha}-\eta_{\beta}+\mathrm{d}\eta_{\alpha\beta } \\
        G_{\alpha\beta\gamma} &\mapsto G_{\alpha\beta\gamma} + \eta_{\alpha\beta }+\eta_{\beta\gamma }+\eta_{\gamma\alpha }
    \end{aligned}
\end{equation}
The $2$-morphisms between $1$-morphisms (in physical words the gauge-of-gauge transformations of the bundle gerbe) are given by collections $(\epsilon_{\alpha })$ of local scalars on each $U_\alpha$ so that
\begin{equation}
    \begin{aligned}
        \eta_{\alpha } &\Mapsto \eta_{\alpha } + \mathrm{d}\epsilon_{\alpha }, \\
        \eta_{\alpha\beta } &\Mapsto \eta_{\alpha\beta }+\epsilon_{\alpha }-\epsilon_{\beta } .
    \end{aligned}
\end{equation}
Pictorially, we can write the coboundaries of abelian bundle gerbes with connection as follows:
\begin{equation}
    \mathbf{H}\big(M,\mathbf{B}^2\mathrm{U}(1)_{\mathrm{conn}}\big) \,\simeq\, \left\{\; \begin{tikzcd}[row sep=scriptsize, column sep=26ex]
    \, M \arrow[r, bend left=60, ""{name=U, below}, "(B_{\alpha }\text{,}\,\Lambda_{\alpha\beta }\text{,}\,G_{\alpha\beta\gamma})"]
    \arrow[r, bend right=60, "(B'_{\alpha }\text{,}\,\Lambda'_{\alpha\beta }\text{,}\,G'_{\alpha\beta\gamma})"', ""{name=D}]
    & \qquad\quad\,\mathbf{B}^2\mathrm{U}(1)_{\mathrm{conn}}
    \arrow[Rightarrow, from=U, to=D, bend left=55, "(\eta'_{\alpha }\text{,}\,\eta'_{\alpha\beta })", ""{name=R, below}] \arrow[Rightarrow, from=U, to=D, bend right=55, "(\eta_{\alpha }\text{,}\,\eta_{\alpha\beta })"', ""{name=L}] \tarrow[from=L, to=R, end anchor={[yshift=0.6ex]}]{r} \arrow["{\small (\epsilon_{\alpha })}", phantom, from=L, to=R, end anchor={[yshift=0.6ex]}, bend left=34]
\end{tikzcd}\right\}
\end{equation}
The \textit{curvature} $H\in\Omega^{3}_\mathrm{cl}(M)$ of a bundle gerbe with connection, i.e. the Kalb-Ramond field strength, is defined by $H|_{U_\alpha}= \di B_\alpha$, which is well-defined since $\di B_\alpha|_{U_\beta} = \di B_\beta|_{U_\alpha}$.

\paragraph{F1-string as higher electric monopole.} 
An F1-string $\phi:\Sigma_2\rightarrow M$ with kinetic action $S_{\mathrm{kin}}(\phi)$ can be naturally interpreted as a \textit{higher electric monopole}, with coupling given by parallel transport
\begin{equation}
    S_{\mathrm{F1}}(\phi) \;=\; S_{\mathrm{kin}}(\phi) \,+ \int_{\Sigma_2} \phi^\ast B.
\end{equation}

\paragraph{NS5-brane as higher magnetic monopole.} 
Recall that a Dirac monopole is a circle bundle of the form $\mathbb{R}^{1}\times(\mathbb{R}^{3}-\{0\}) \rightarrow \mathbf{B}\mathrm{U}(1)$ with non-trivial first Chern class in $\mathrm{H}^2(\mathbb{R}^{1}\times(\mathbb{R}^{3}-\{0\},\mathbb{Z})\cong \mathbb{Z}$. Here, $\mathbb{R}^{1}$ can be seen as a magnetically charged world-line, i.e. a magnetic monopole.
In direct analogy with this, a \textit{higher Dirac monopole} can be constructed as a bundle gerbe of the form
\begin{equation}
    \mathbb{R}^{1,5} \times (\mathbb{R}^4-\{0\}) \longrightarrow \mathbf{B}^2\mathrm{U}(1),
\end{equation}
with non-trivial Dixmier-Douady class. Here, $\mathbb{R}^{1,5}$ can be seen as the world-volume of an NS5-brane magnetically charged by the Kalb-Ramond field.
This spacetime is homotopy equivalent to the $3$-sphere, so the $3$rd cohomology group will be isomorphic to $\mathrm{H}^3(S^3,\mathbb{Z})\cong\mathbb{Z}$. Thus, the Dixmier-Douady number of any such bundle will be an integer
\begin{equation}
    \frac{1}{4\pi^2}\int_{S^3} H = m\in\mathbb{Z},
\end{equation}
which could be called higher magnetic charge, or $H$-charge.

\subsection{String bundles and heterotic string theory}

Heterotic string theory is a type of string theory where the Kalb-Ramond field is twisted by the appearance of gauge fields with structure group either $G = SO(32)$ or $G=E_8\times E_8$. Remarkably, the gauge sector of heterotic string theory can be still formulated as a higher gauge theory with a non-abelian principal $\infty$-bundle.
The smooth $\infty$-group which encodes the higher gauge theory of heterotic string theory is denoted by $\mathrm{String}_{\mathrm{het}\!}^{\mathbf{c}_2}(1,9)$ and defined by the pullback diagram
\begin{equation}\begin{tikzcd}[row sep=14ex, column sep=7.0ex]
\mathbf{B}\mathrm{String}_{\mathrm{het}\!}^{\mathbf{c}_2}(1,9) \arrow[d, "\mathrm{hofib}\!\,\left(\frac{1}{2}\mathbf{p}_1-\mathbf{c}_2\right)"']\arrow[r] &\ast \arrow[d]  \\
\mathbf{B}\big(\mathrm{Spin}(1,9)\times G\big)\arrow[r, "\frac{1}{2}\mathbf{p}_1-\mathbf{c}_2"] &\mathbf{B}^3\mathrm{U}(1),
\end{tikzcd}\end{equation}
where 
\begin{itemize}
    \item the map $\frac{1}{2}\mathbf{p}_1: \mathbf{B}\mathrm{Spin}(1,9) \rightarrow \mathbf{B}^3\mathrm{U}(1)$ is the smooth refinement of the $1$st fractional Pontryagin class of the frame bundle $FM\twoheadrightarrow M$, that is given by a non-abelian cocycle $M\rightarrow \mathbf{B}\mathrm{Spin}(1,9)$;
    \item the map $\mathbf{c}_2: \mathbf{B}G \rightarrow \mathbf{B}^3\mathrm{U}(1)$ is the smooth refinement of the second Chern class of a principal $G$-bundle $P\twoheadrightarrow M$, that is given by a cocycle $M\rightarrow \mathbf{B}G$.
\end{itemize}
Such a definition makes any principal $\mathrm{String}_{\mathrm{het}\!}^{\mathbf{c}_2}(1,9)$-bundle $P_{\mathrm{het}}\twoheadrightarrow M$ a twisted bundle gerbe over an ordinary $\mathrm{Spin}(1,9)\times G$-bundle, i.e. a pullback diagram of the following form:
\begin{equation}\begin{tikzcd}[row sep=11ex, column sep=2ex]
P_{\mathrm{het}}\arrow[d]\arrow[r] &\ast \arrow[d] \\ 
FM\!\times_{M}\!P  \arrow[d]\arrow[r] &\mathbf{B}^2\mathrm{U}(1) \arrow[d]\arrow[r] & \ast \arrow[d]  \\
M\arrow[r, "g"] & \mathbf{B}\mathrm{String}_{\mathrm{het}\!}^{\mathbf{c}_2}(1,9) \arrow[r] &\mathbf{B}(\mathrm{Spin}(1,9)\times G).
\end{tikzcd}\end{equation}

\paragraph{Heterotic string theory.}
Locally on any open patch $U\subset M$ of spacetime, the connection of the $\mathrm{String}_{\mathrm{het}\!}^{\mathbf{c}_2}(1,9)$-bundle is given by a triple $(\omega,A,B)$, where
\begin{equation}
\begin{aligned}
     \omega \,&\in\, \Omega^1(U,\mathfrak{spin}(1,9)) && \;\text{(spin connection)},\\
     A\,&\in\, \Omega^1(U,\mathfrak{g}) && \;\text{(gauge field)}, \\
     B\,&\in\, \Omega^2(U) && \;\text{(Kalb-Ramond field)}.
\end{aligned}
\end{equation}
The curvature of a $\mathrm{String}_{\mathrm{het}\!}^{\mathbf{c}_2}(d)$-bundle will be given by differential forms
\begin{equation}
    \begin{aligned}
    R_{b}^{\;\,a} \;&=\; \di  \omega_{b}^{\;\,a} + \omega_{c}^{\;\,a} \wedge  \omega_{b}^{\;\,c} && \;\text{(curvature)}, \\
    F \;&=\; \di  A + [ A \,\overset{\wedge}{,}\,  A ]_{\mathfrak{g}} && \;\text{(gauge field strength)}, \\
    H \;&=\; \di B - \mathrm{cs}_3(A) +  \mathrm{cs}_3(\omega) && \;\text{(Kalb-Ramond field strength)},
    \end{aligned}
\end{equation}
where $\mathrm{cs}_3(A_{\alpha })$ and $\mathrm{cs}_3(\omega_{\alpha })$ are the Chern-Simons differential $3$-forms of the two connections.

\begin{equation}
    \begin{aligned}
    \di R_{b}^{\;\,a} + \omega_{c}^{\;\,a} \wedge  R_{b}^{\;\,c} \;&=\; 0, &&\\
    \di F + [ A \,\overset{\wedge}{,}\,  F ]_{\mathfrak{g}} \;&=\; 0 && \text{(Bianchi identities)}\\
    \di H + \langle F \,\overset{\wedge}{,}\,F \rangle_\mathfrak{g} - R_{b}^{\;\,a} \wedge R_{a}^{\;\,b} \;&=\; 0. &&
    \end{aligned}
\end{equation}

\subsection{String dualities on $\infty$-bundles}

Recent advances in String Theory have shed new light on string dualities through the lens of principal $\infty$-bundles, e.g. see \cite{BelHulMin07, Hull14, DesSae18, NikWal18, DesSae19, Alf19, Alfonsi:2021ymc, Alfonsi:2021uwh, Kim:2022opr}. 
To convey why this has proved to successful, let us briefly sketch a picture of topological T-duality and point at the natural appearance of higher geometry.

\paragraph{Topological T-duality.}
First, consider two spacetime manifolds $M,\widetilde{M}$. Say that they are both $T^n$-bundles ${\pi}:M\rightarrow N$ and ${\widetilde{\pi}}:\widetilde{M}\rightarrow N$ over a common base manifold $N$, with first Chern classes respectively given by $\mathrm{c}_1(M)\in \mathrm{H}^2(N,\mathbb{Z})^n$ and $\widetilde{\mathrm{c}}_1(\widetilde{M})\in \mathrm{H}^2(N,\mathbb{Z})^n$. 
Now, consider a pair of bundle gerbes $\Pi:P\rightarrow{}M$ and $\widetilde{\Pi}:\widetilde{P}\rightarrow{}\widetilde{M}$, encoding the topological data of two Kalb-Ramond fields, respectively on $M$ and $\widetilde{M}$, with Dixmier-Douady classes of the form
\begin{equation}
    [H] \,=\, \bigg[ \sum_{i=1}^n h_i\otimes \widetilde{\mathrm{c}}_1(\widetilde{M})^i\bigg]\in \mathrm{H}^3(M,\mathbb{Z}), \qquad [\widetilde{H}] \,=\, \bigg[ \sum_{i=1}^n \widetilde{h}^i\otimes \mathrm{c}_1(M)_i\bigg]\in \mathrm{H}^3(\widetilde{M},\mathbb{Z}),
\end{equation}
where $h_i$ and $\widetilde{h}^i$ are respectively the generators of the cohomology of the fibres $T^n$ and $\widetilde{T}^n$.
Then, the bundle gerbes $\Pi:P\rightarrow{}M$ and $\widetilde{\Pi}:\widetilde{P}\rightarrow{}\widetilde{M}$ are geometric T-dual if there exists an isomorphism
\begin{equation}
    \begin{tikzcd}[row sep={9.5ex,between origins}, column sep={10.5ex,between origins}]
    & P\times_{N} \! \widetilde{M}\arrow[rr, "\simeq"', "\mathrm{T\text{-}duality}"]\arrow[dr, "\Pi"']\arrow[dl, "\widetilde{\pi}"] & & M\times_{N} \!\widetilde{P}\arrow[dr, "\pi"']\arrow[dl, "\widetilde{\Pi}"] \\
    P\arrow[dr, "\Pi"'] & & M\times_{N} \!\widetilde{M}\arrow[dr, "\pi"']\arrow[dl, "\widetilde{\pi}"] & & [-2.5em]\widetilde{P}\arrow[dl, "\widetilde{\Pi}"] \\
    & M\arrow[dr, "\pi"'] & & \widetilde{M}\arrow[dl, "\widetilde{\pi}"] & \\
    & & N & &
    \end{tikzcd}
\end{equation}
such that it satisfies the Poincar\'e condition: that is, that at any point $x\in N$ of the base manifold we must have
\begin{equation}
    \big[\mathrm{T\text{-}duality}|_x\big] \,=\, \bigg[ \sum_{i=1}^n h_i \smile \widetilde{h}^i \bigg] \;\in\; \frac{\mathrm{H}^2(T^n \times \widetilde{T}^n,\mathbb{Z})}{\!\mathrm{im}(\pi^\ast|_x)\oplus\mathrm{im}(\widetilde{\pi}^\ast|_x)},
\end{equation}
where we used the fact that an isomorphism of bundle gerbes is equivalently a $\mathrm{U}(1)$-bundle on its base manifold.
The wish to differentially refine this picture suggests that a full understanding of T-duality needs to happen in higher geometry.

\paragraph{Non-commutative and non-associative geometry.}
A phenomenon intimately related to string dualities is the appearance of non-associative geometry in open string theory. This stringy-geometric feature is understood to be linked, first, to the non-geometric fluxes typically produced by T-duality \cite{SN1, SN2, SN4} and, second, to higher differential geometry \cite{SN3, SN5, SN6, SN7, Sza18}.

\paragraph{Tensor hierarchy and gauged supergravity.}
The idea of \textit{tensor hierarchy} was introduced by \cite{deWit:2008ta} as a dimensional reduction of duality-covariant string theory. Then, it was further developed by \cite{Kotov:2018vcz, Hohm19DFT, Hohm19, Hohm19x, Ced20a, Ced20b} as a higher gauge theory.

\subsection{Courant algebroids and generalized geometry}
A {Courant algebroid} $E\twoheadrightarrow M$ is a vector bundle given by a short exact sequence of vector bundles
\begin{equation}
    \begin{tikzcd}[row sep=5ex, column sep=6ex] 
    T^\ast M \arrow[r, hook, "\rho^\ast"] & E \arrow[r, two heads, "\rho"] & TM,
\end{tikzcd}
\end{equation}
where $\rho:E\twoheadrightarrow TM$ is known as anchor map and whose space of sections is equipped with a bracket structure generalizing Lie algebroids (see \cite{Gua11}).
In this section, we will show that Courant algebroids, which are the main object studied by generalized geometry, can be seen as higher Atiyah algebroids for principal $\infty$-bundles.

Given any map $g:X\rightarrow Y$ in $\mathbf{H}$, it is possible to construct its automorphism $\infty$-groupoid $\Aut (g)$.
If $P\rightarrow M$ is a principal $\infty$-bundle given by the homotopy fiber of a map $g:M\rightarrow\mathbf{B}G$, then $\Aut (g)$ can be seen as the $\infty$-group of bundle automorphisms, i.e. automorphisms of the stack $P$ preserving the principal bundle structure
This $\infty$-group will sit at the center of a short exact sequence of $\infty$-groups
\begin{equation}\label{eq:autdef}
    1\longrightarrow\Omega_g\mathbf{H}(M,\mathbf{B}G)\longrightarrow \Aut (g) \longrightarrow \mathrm{Diff}(M)\longrightarrow 1.
\end{equation}
The \textit{Atiyah $L_\infty$-algebra} of an $\infty$-bundle was defined by \cite{FSS16} as the Lie differentiation of its automorphism $\infty$-groupoid, i.e. 
$\mathfrak{at}(P) \coloneqq  \mathrm{Lie}(\Aut_{\mathbf{H}_/}(g))$.
This $L_\infty$-algebra encodes the infinitesimal symmetries of the principal structure. By differentiating \eqref{eq:autdef} we have the short exact sequence of $L_\infty$-algebras
\begin{equation}
    0\longrightarrow\mathrm{Lie}\big(\Omega_g\mathbf{H}(M,\mathbf{B}G)\big)\longrightarrow \mathfrak{at}(P) \longrightarrow \mathfrak{X}(M)\longrightarrow 0.
\end{equation}
For example, if $P\rightarrow M$ is an ordinary principal $G$-bundle, we have $\Omega_g\mathbf{H}(M,\mathbf{B}G)\simeq \Gamma\big(M,P\times_G G\big)$.
Thus, we get the short exact sequence
\begin{equation}
    0\longrightarrow\Gamma(M,\mathfrak{g}_P)\longrightarrow \mathfrak{at}(P) \longrightarrow \mathfrak{X}(M)\longrightarrow 0,
\end{equation}
where $\mathfrak{g}_P\coloneqq P\times_G\mathfrak{g}$ is the {adjoint bundle} and $\mathfrak{X}(M)$ is the Lie algebra of vector fields on $M$. In the ordinary case, we recover the algebra of sections $\mathfrak{at}(P)\cong \Gamma(M,TP/G)$ of the usual Atiyah algebroid.

Now, let $P\rightarrow M$ be a bundle gerbe with partial connection data corresponding to a map $(\Lambda,G):M\rightarrow\mathbf{B}(\BU)$. Then, it is possible to prove that there is an equivalence of $\infty$-groups $\Omega_{(\Lambda,G)}\mathbf{H}(M,\mathbf{B}(\BU)) \simeq \mathbf{H}\big(M,\BU\big)$.
As explained by \cite{Col11, Rog11}, we get that the Atiyah $L_\infty$-algebra $\mathfrak{at}(P)$ coincides with the so-called \textit{Courant $2$-algebra} of sections of a Courant algebroid.
Locally, on a patch $U\subset M$, this reduces the familiar complex of infinitesimal gauge transformations of the Kalb-Ramond field
\begin{equation}
    \mathfrak{at}(P)|_U \;\simeq\; \Big( \Coo(U)\,\xrightarrow{\;\mathrm{d}\;}\,\Gamma(U,TU\oplus T^\ast U) \Big).
\end{equation}

\section{Higher geometric quantization}

Higher geometric quantization \cite{Rog11,SaSza11x,Rog13,SaSza13, FRS18, Fiorenza:2013jz, FSS16, BSS16,BS16,Sza19} is a mathematical framework for quantizing a classical field theory which generalizes ordinary geometric quantization. See \cite{Bunk:2021quu} for a detailed review.

Recall that ordinary geometric quantization is a well-established method for constructing a non-perturbative quantization of the phase space of a classical mechanical system, i.e. a symplectic manifold $(M,\omega)$. 
This is achieved by the construction of a prequantum bundle $(P,\nabla)$ on the symplectic manifold $(M,\omega)$, which is a principal $\mathrm{U}(1)$-bundle $P\twoheadrightarrow M$ equipped with a connection $\nabla$, whose curvature is $F_\nabla=\omega\in\Omega^2_{\mathrm{cl}}(M)$. The prequantum Hilbert space is constructed as the vector space $\mathfrak{H}_{\mathrm{pre}} \coloneqq \Gamma(M, E)$ of sections of the $\mathbb{C}$-associated bundle $E \coloneqq  P\times_{\mathrm{U}(1)}\mathbb{C}$.

Now, \textit{higher geometric quantization} generalizes ordinary geometric quantization in two orthogonal directions:
\begin{itemize}
    \item[(\textit{a})] the ordinary prequantum $\mathrm{U}(1)$-bundle is generalized by a prequantum $\B^{n}\mathrm{U}(1)$-bundle, i.e. a bundle $n$-gerbe, for $n\in\mathbb{N}$;
    \item[(\textit{b})] the ordinary phase space $M$ is generalized by a smooth stack, as argued by \cite{Sev01} and further developed by \cite{FRS18, Fiorenza:2013jz, FSS16}.
\end{itemize}

\subsection{Higher symplectic (or $n$-plectic) geometry}\label{sec:n_plectic}

To proceed in the (\textit{a}) direction, we need to introduce a higher generalization of symplectic geometry.
An \textit{$n$-plectic manifold} $(M,\omega)$ is defined as a smooth manifold $M$ equipped with a closed $(n+1)$-form $\omega\in\Omega^{n+1}_{\mathrm{cl}}(M)$, satisfying a weak non-degeneracy condition (see \cite{Rog11,Rog13}).

\paragraph{Higher Poisson algebras.}
$n$-plectic manifolds come with a natural notion of \textit{Hamiltonian forms}.
An ($n-1$)-form is called Hamiltonian if there exists a vector $V_\xi\in\mathfrak{X}(M)$, which is called \textit{Hamiltonian vector}, such that $\iota_{V_\xi}\omega= \di \xi$. Let us denote by $\Omega^{n-1}_{\mathrm{Ham}}(M,\omega)$ the space of Hamiltonian $(n-1)$-forms.
If we consider the complex
\begin{equation}
    \Omega^0(M) \xrightarrow{\,\di\,} \Omega^1(M)  \xrightarrow{\,\di\,} \dots  \xrightarrow{\,\di\,} \Omega^{n-2}(M)  \xrightarrow{\,\di\,} \Omega^{n-1}_{\mathrm{Ham}}(M,\omega) 
\end{equation}
concentrated in degrees from $-n$ to $0$,
this can be equipped with the $L_\infty$-algebra structure:
\begin{equation}
    \ell_k(\xi_1,\dots,\xi_k ) \,=\, \begin{cases}-(-1)^{\left(\substack{k+1\\2}\right)} \iota_{V_{\xi_1}\wedge \cdots \wedge V_{\xi_k}}\omega, & \text{if }|\xi_1|=\dots=|\xi_k|=0 \\ 0, &\text{otherwise.}\end{cases}
\end{equation}
This is the \textit{higher Poisson algebra} of observables on the $n$-plectic manifold $(M,\omega)$.

\subsection{Higher prequantum Hilbert spaces}

Let $(M,\omega)$ be an $n$-plectic manifold.
A \textit{prequantum $\infty$-bundle} on an $n$-plectic manifold $(M,\omega)$ is a bundle $n$-gerbe $P\rightarrow M$ whose curvature is $F_\nabla=\omega\in\Omega^{n+1}_{\mathrm{cl}}(M)$.
The prequantum $n$-Hilbert space is then defined by the $\infty$-groupoid $\mathfrak{H}_{\mathrm{pre}} \coloneqq  \Gamma(M,E)$ of sections of the $V$-associated $\infty$-bundle 
\begin{equation}
    E \,\coloneqq\,  P\times_{\mathbf{B}^n\mathrm{U}(1)}V,
\end{equation}
where $V$ is any smooth stack equipped with a $\mathbf{B}^n\mathrm{U}(1)$-action. 

Now, let us focus on the case $n=2$. First, notice that there exist a canonical inclusion of unitary groups $\mathrm{U}(N) \hookrightarrow \mathrm{U}(N+1)$ for any positive natural number $N$.
So, we can define the moduli stack $\mathbf{B}\mathrm{U} = \mathrm{hocolim}_N\mathbf{B}\mathrm{U}(N)$ and, given a prequantum bundle gerbe $P$, its $\mathbf{B}\mathrm{U}$-associated bundle $E = P\times_{\mathbf{B}\mathrm{U}(1)}\mathbf{B}U$.
Thus, the prequantum Hilbert space can be identified with the $\infty$-groupoid
\begin{equation}
    \mathfrak{H}_{\mathrm{pre}} \;=\; \Gamma\big(M,\, P\times_{\mathbf{B}\mathrm{U}(1)}\mathbf{B}\mathrm{U} \big),
\end{equation}
whose objects are $\mathrm{U}(N)$-bundles twisted by $P$ on the base manifold $M$.
For a construction of a higher Hilbert spaces, equipped with a categorified notion of Hilbert product, see \cite{BSS16,BS16}.

\paragraph{Example: closed string.} An interesting example is the following. Consider a prequantum bundle gerbe $P\rightarrow M$ given by a cocycle $g:M\rightarrow \mathbf{B}^2\mathrm{U}(1)_\mathrm{conn}$. Such a bundle gerbe can be transgressed to a $\mathrm{U}(1)$-bundle $\hat{P}$ on the loop space $\mathcal{L}M$ of the original smooth manifold by 
\begin{equation}
    \mathcal{L}M \,\simeq\, [S^1,M] \;\xrightarrow{\;[S^1\!,\,g]\;}\; [S^1,\mathbf{B}^2\mathrm{U}(1)_\mathrm{conn}] \;\xrightarrow{\;\int_{S^1}\;}\; \mathbf{B}\mathrm{U}(1)_\mathrm{conn}.
\end{equation}
This means that the $\mathbf{B}\mathrm{U}$-associated bundle $P\times_{\mathbf{B}\mathrm{U}(1)}\mathbf{B}\mathrm{U} \longtwoheadrightarrow M$, given by a twisted cocycle
\begin{equation}
    M\;\longrightarrow \; \mathbf{B}\mathrm{U}/\!/\mathbf{B}\mathrm{U}(1),
\end{equation}
will be transgressed to an ordinary $\mathbb{C}$-associated bundle $\hat{P}\times_{\mathrm{U}(1)}\mathbb{C}\longtwoheadrightarrow\mathcal{L}M$, given by the transgressed cocycle $[S^1,M]\rightarrow \mathbb{C}/\!/\mathrm{U}(1)$ on the loop space $\mathcal{L}M\simeq[S^1,M]$ of the smooth manifold $M$. 
See \cite{SaSza11x,SaSza13} for discussion of higher geometric quantization and loop spaces.

\paragraph{Example: Chern-Simons theory.} 
As mentioned above, It is possible to construct a prequantum $\infty$-bundle, more generally, on a smooth stack \cite{Sev01, Fiorenza:2013jz, FSS16} and not necessarily on a manifold.
A relevant example \cite{FRS18, Fiorenza:2013jz, FSS16} is provided by the construction of a prequantum bundle $2$-gerbes on the moduli stack $\B G$ of principal $G$-bundles, for some Lie group $G$. Such a prequantum bundle
\begin{equation}
    \mathbf{c}_2:\, \mathbf{B}G_\mathrm{conn} \;\longrightarrow\; \mathbf{B}^{3}\mathrm{U}(1)_\mathrm{conn}
\end{equation}
is given by a smooth differential refinement of a $2$nd Chern class, that is, a characteristic class $c_2\in\mathrm{H}^4(BG,\mathbb{Z})$ on the topological classifying space $BG$.
If we transgress such a prequantum $\infty$-bundle to a smooth $2$-manifold $\Sigma_2$, we obtain an ordinary prequantum bundle $[\Sigma_2,\mathbf{B}G_\mathrm{conn}] \rightarrow \mathbf{B}\mathrm{U}(1)_\mathrm{conn}$.
This setting is very closely related to the usual quantization of $3$d Chern-Simons theory. To see this, notice that we can also transgress the original prequantum $\infty$-bundle on a smooth $3$-manifold $\Sigma_3$, so that we obtain the map
\begin{equation}
\begin{aligned}
    Z:\,[\Sigma_3,\mathbf{B}G_\mathrm{conn}] \;&\longrightarrow\; \mathrm{U}(1) \\
     A \;&\longmapsto\;  \exp\Big(i\!\int_{\Sigma_3}\!\mathrm{CS}_3(A)\Big).
\end{aligned}
\end{equation}
Other relent examples of higher geometric quantization on smooth stacks include Poisson smooth groupoids (see \cite{Sev01, FSS16}).

\section{Hypothesis H}

In this section, we will provide a brief introduction to Hypothesis H \cite{FSS19coho, FSS19xx, Fiorenza:2019usl}.
This is a promising conjecture in mathematical physics stating that the charge quantization of the $C$-field of M-theory must take value in the non-abelian cohomology theory known as J-twisted cohomotopy, in the same way that D-brane charges are quantized in twisted K-theory (table \ref{tab:cohomology}).

\paragraph{Towards cohomotopy.} The equations of motion and Bianchi identities of the gauge field sector of $11$d supergravity on a spacetime manifold $M$ can be entirely given by
\begin{equation}\label{eq:sugra}
\begin{aligned}
    \di F_4 \,&=\, 0,\\
    \di F_7 + F_4\wedge F_4 \,&=\, 0,
\end{aligned}
\end{equation}
where $F_4\in\Omega^4(M)$ is the field strength of the $C$-field and $F_7\in\Omega^7(M)$ is the one of its the magnetic dual.
The first line of equation \eqref{eq:sugra} could be understood as the Bianchi identity of a bundle $2$-gerbe, i.e. a principal $\B^2\mathrm{U}(1)$-bundle, but the second line is not as easy to treat.
In \cite{FSS15x}, it was observed that the curvature of a bundle $5$-gerbe twisted by a bundle $2$-gerbe, as in the pullback diagram
\begin{equation}
\begin{tikzcd}[row sep={16ex,between origins}, column sep={17ex,between origins}]
P_{\mathrm{M5}}\arrow[r]\arrow[d, "\Pi_{\mathrm{M5}}"'] & \ast \arrow[d] & \\
P_{\mathrm{M2}}\arrow[r, "g_{\mathrm{M5}}"]\arrow[d, "\Pi_{\mathrm{M2}}"'] & \mathbf{B}^6\mathrm{U}(1) \arrow[r]\arrow[d] & \ast \arrow[d]\\
M \arrow[r, "g_{\mathrm{M2/5}}"]\arrow[rr, "g_{\mathrm{M2}}", bend right=30]& {\mathbf{B}^6\mathrm{U}(1)/\!/\mathbf{B}^2\mathrm{U}(1)} \arrow[r] & \mathbf{B}^3\mathrm{U}(1),
\end{tikzcd}
\end{equation}
satisfies the equations \eqref{eq:sugra} of the $C$-field.
However, as pointed out in \cite{FSS15x}, the twisted cocycle $g_{\mathrm{M2/5}}$ in the diagram above can be further generalized to a cocycle $M\rightarrow S^4$ valued in the rationalized $4$-sphere. In fact, the Sullivan model of the $4$-sphere is given by the dg-algebra 
\begin{equation}\label{sphere}
    \mathrm{CE}(\mathfrak{l}S^4)\;=\; \mathbb{R}[f_4,f_7]/(\di f_4=0,\;\di f_7 + f_4\wedge f_4=0),
\end{equation}
where $f_4$ and $f_7$ are generators respectively in degree $4$ and $7$. Thus, a cocycle valued in the rationalized $4$-sphere fully reproduces the equations of motion of the gauge field sector of $11$d supergravity. This fact suggests that the gauge sector of supergravity could be charge-quantized in a generalized cohomology theory with coefficients in the homotopy types of $n$-spheres.

\paragraph{Twisted cohomotopy.} Let $\{A_n\}_{n\in\mathbb{N}}$ be a spectrum, i.e. a sequence of topological spaces which are the loop spaces of each other $A_n=\Omega A_{n+1}$. Then, a \textit{generalized cohomology theory} is given by $\mathrm{H}(X;A_n) \coloneqq \mathrm{Hom}(X,A_n)_{/\sim}$ that is the set of homotopy classes of continuous maps $X\rightarrow A_n$. 
If we consider the spectrum $\{S^n\}_{n\in\mathbb{N}}$ of spheres, we obtain the generalized cohomology theory known as \textit{cohomotopy}, and the $n$th cohomotopy of a space $X$ is given by the set $\mathrm{H}(X;S^{n})$.

Now, let $\tau$ be a Riemannian structure on a smooth $d$-manifold $M_d$. The \textit{J-twisted cohomotopy} on $M_d$ in degree $d-1$ is given by the set
\begin{equation}
    \mathrm{H}^\tau(M_d;S^{d-1}) \;=\; \left\{\begin{tikzcd}[row sep={10ex,between origins}, column sep={18ex,between origins}]
    M_d \arrow[r]\arrow[rd, "\tau"] & {S^{d-1}/\!/\mathrm{O}(d)}  \arrow[d] \\
    & B\mathrm{O}(d)
    \end{tikzcd} \right\}_{/\sim}
\end{equation}
\textit{Hypothesis H} proposes that the $C$-field $4$-form and $7$-form fluxes of M-theory are charge-quantized in J-twisted cohomotopy cohomology theory. 
For instance, if we consider the case of a $\mathrm{Sp}(2)\times \mathrm{Sp}(1)$-structure $\tau$ on a spin $8$-manifold $M_8$, one recovers the $\tau$-twisted cohomotopy
\begin{equation}
    \mathrm{H}^\tau(M_8;S^{4}) \;=\; \left\{\begin{tikzcd}[row sep={10ex,between origins}, column sep={18ex,between origins}]
    M_8 \arrow[r]\arrow[rd, "\tau"] & {S^4/\!/\mathrm{Sp}(2)\times \mathrm{Sp}(1)}  \arrow[d] \\
    & B(\mathrm{Sp}(2)\cdot \mathrm{Sp}(1))
    \end{tikzcd} \right\}_{/\sim}
\end{equation}
Given Hypothesis H, the $C$-field $4$- and $7$-form fluxes of M-theory must in the image of the non-abelian Chern character map from J-twisted cohomotopy theory.
For instance, if we differentially refine the case above of the spin $8$-manifold $M_d$ with an affine connection $\nabla$, one has that the de~Rham cohomology is given by the set
\begin{equation*}
    \mathrm{H}^{\tau_\dR}_\dR(M_8;S^4) \,=\, \left\{\,\begin{matrix}F_4\in\Omega^4(M_8) \\[0.4ex]  F_7\in\Omega^7(M_8)\end{matrix}\;\,\left|\,\;\begin{matrix}\qquad\qquad\qquad\qquad\qquad\qquad\qquad\qquad\qquad\quad \di F_4\!\!&\!\!=\,0\\[0.4ex] \di F_7 + \big(F_4 - \frac{1}{4}p_1(\nabla)\big)\wedge \big(F_4- \frac{1}{4}p_1(\nabla)\big) + \chi_8(\nabla)\!\!&\!\!=\,0\end{matrix}\right.\,\right\}_{/\sim}
\end{equation*}
where $p_1(\nabla)$ is the $1$st Pontrjagin form and $\chi_8(\nabla)$ is the Euler form of the connection. Remarkably, on 8-manifolds, this charge quantization
recovers the expected anomaly cancellation conditions of M-theory.
One can also recover expected phenomena in microscopic D-brane physics \cite{Sati:2019nli}.

\begin{table}[h!]\begin{center}
\begin{center}
 \begin{center}
\begin{tabular}{||c c c||} 
 \hline
 Cohomology theory & on $M$ & Charge quantisation \\ [0.5ex] 
 \hline\hline
 $2$nd integral cohomology & $\mathrm{H}^2(M,\mathbb{Z})$ & magnetic charge \\ 
 \hline
 $3$rd integral cohomology & $\mathrm{H}^3(M,\mathbb{Z})$ & fundamental string charge \\
 \hline
 $\mathrm{H}^3$-twisted K-theory & $\mathrm{H}^\tau(M;\mathrm{KU})$ & D-branes charges \\
 \hline
 J-twisted cohomotopy & $\mathrm{H}^\tau(M;S^4)$ & M-branes charges \\ [1ex] 
 \hline
\end{tabular}
\end{center}
\end{center}
\caption{\label{tab:cohomology}Proposals of charge quantization in theoretical physics.}\vspace{-0.2cm}
\end{center}\end{table}

\section{Derived geometry and Batalin-Vilkovisky formalism}

We will provide a brief overview on Batalin-Vilkovisky (BV-)theory. For a more detailed introduction to BV-theory, we point to the article \cite{cattaneo2023bv}.
In the settings of several broader programmes, there exist various different approaches to BV-theory,  including:
\begin{itemize}
    \item $L_\infty$\textit{-algebras} approach (see \cite{Saem18bv, Saem19bv, Doubek_2019, Jurco:2019yfd, Jurco:2020yyu}), where the algebra of classical observables is given by a Poisson dg-Lie algebra of functions on an $L_\infty$-algebra equipped with a $(-1)$-shifted symplectic form. 
    \item \textit{Factorization Algebras} approach (see \cite{Costello2011RenormalizationAE, FactI, FactII}), where the algebra of classical observables is given by the $\mathbb{P}_0$-algebra of functions on a $(-1)$-shifted symplectic pointed formal moduli problem, which is also sheaved on spacetime.
    \item \textit{Perturbative Algebraic Quantum Field Theory} -- shortened as pAQFT -- (see  \cite{Rejzner:2011jsa, Fredenhagen_2012, Fredenhagen:2011an, Rejzner:2016hdj, Benini:2018oeh, Benini:2019uge, Benini_2019,  hawkins2020star, Rejzner:2020bsc, Rejzner:2020xid}), where the algebra of observables is given by a net of Poisson dg-$\ast$-algebras.
\end{itemize}
Despite their differences, these approaches can be connected. 
In fact, $L_\infty$-algebras are closely related to the formal moduli problems appearing in the factorization algebras approach. In particular, they both give rise to an $L_\infty$-structure on the space of classical observables, as seen by \cite{Saem19bv, FactII}.
Moreover, the factorization algebra and pAQFT formalisms are understood to be intimately related, via a correspondence which was delineated and explored by \cite{Gwilliam:2017ses, Benini:2019ujs}.

\subsection{BV-formalism in terms of $L_\infty$-algebras.}

Consider a BRST (local)\footnote{One must be careful when dealing with infinite-dimensional $L_\infty$-algebras. In this article we ignore these subtleties, but, as explained by \cite{FactI}, we should work with local $L_\infty$-algebras (whose spaces of sections are bornological vector spaces).} encoding the kinematics of a classical field theory.
Such an $L_\infty$-structure can be dually given by its Chevalley-Eilenberg dg-algebra $\CE(\mathfrak{L})$, which is of the form 
\begin{equation}
    \CE(\mathfrak{L}) \;=\; \big(\mathrm{Sym}\,\mathfrak{L}^\vee[-1],\;\di_{\CE(\mathfrak{L})}\big).
\end{equation}
This is precisely the \textit{BRST complex} in physics. The second ingredient of a BV-theory is the action functional of our field theory, which can be regarded as an element $S \in \CE(\mathfrak{L})$.

Consider the graded vector space $\mathfrak{L}[1]$, which is the graded manifold with the property that $\Coo(\mathfrak{L}[1]) = \mathrm{Sym}\,\mathfrak{L}^\vee[-1]$.
Then, the machinery of BV-theory instructs us to take the $(-1)$-shifted cotangent bundle of such a graded vector space, that is
\begin{equation}
    T^\vee[-1]\mathfrak{L}[1] \;=\; (\mathfrak{L} \oplus \mathfrak{L}^\vee[-3])[1].
\end{equation}
Similarly to ordinary cotangent bundles, a $(-1)$-shifted cotangent bundle comes equipped with a natural $(-1)$-shifted Poisson bracket $\{-,-\}$.
The goal of this recipe is to equip our new graded vector space $\mathfrak{L} \oplus \mathfrak{L}^\vee[-3]$ with the structure of a certain $L_\infty$-algebra extending our BRST $L_\infty$-algebra $\mathfrak{L}$. 
To do that, we can define the \textit{classical BV-action} $S_\BV\in\mathrm{Sym}(\mathfrak{L}\oplus \mathfrak{L}^\vee[-3])^\vee[-1]$ by the sum
\begin{equation}
    S_\BV \;=\; S + S_{\mathrm{BRST}},
\end{equation}
where $S \in \CE(\mathfrak{L})$ is the original action of the theory and $S_{\mathrm{BRST}}\coloneqq\widehat{\di_{\CE(\mathfrak{L})}}$ is the cotangent lift of the original Chevalley-Eilenberg differential $\di_{\CE(\mathfrak{L})}$.
The classical BV-action satisfies the so-called \textit{classical master equation}: 
\begin{equation}
    \{S_{\BV},S_{\BV}\} \;=\; 0.
\end{equation}
We can define the BV-differential by $Q_\BV \coloneqq \{S_{\BV},-\}$, so that the classical master equation is equivalent to $Q_\BV^2 = 0$. 
Moreover, we notice that $\mathrm{Sym}(\mathfrak{L}\oplus \mathfrak{L}^\vee[-3])^\vee[-1] = \mathrm{Sym}(\mathfrak{L}^\vee[-1]\oplus \mathfrak{L}[2])$.
Thus, we have all we need to define the following Chevalley-Eilenberg cochain: 
\begin{equation}
    \CE\big(\mathfrak{Crit}(S)\big) \;\coloneqq\; \big(\mathrm{Sym}(\mathfrak{L}^\vee[-1]\oplus \mathfrak{L}[2]),\;\, Q_\BV = \{S_\BV,-\} \big).
\end{equation}
This can be dually interpreted  as a (local) $L_\infty$-algebra $\mathfrak{Crit}(S)$, whose underlying graded vector space is $T^\vee[-1]\mathfrak{L}[1]$, as wanted.
The $(-1)$-shifted Poisson bracket provides $\mathfrak{Crit}(S)$ with a \textit{cyclic structure}.
The Chevalley-Eilenberg cochain $\CE\big(\mathfrak{Crit}(S)\big)$ is known as \textit{BV-complex} in physics literature.

We can compute the \textit{minimal model} of $\mathfrak{Crit}(S)$ by transferring by deformation retract its $L_\infty$-algebra structure to the cohomology $\mathrm{H}^\bullet\big(\mathfrak{Crit}(S)\big)$. In \cite{Jurco:2019yfd, Macrelli:2019afx, Jurco:2020yyu} it was shown that the brackets of the minimal model correspond to scattering amplitudes (see table \ref{tab:Loo}).

\begin{table}[h!]\begin{center}
\begin{center}
\begin{tabular}{||c c||} 
 \hline
 $L_\infty$-algebras & QFT \\ [0.5ex] 
 \hline\hline
 cyclic $L_\infty$-algebra & perturbative field theory \\ 
 \hline
 \shortstack{minimal model of\\cyclic $L_\infty$-algebra} &  \shortstack{non-trivial part of \\classical S-matrix} \\
 \hline
 \shortstack{{minimal model of }\\{quantum cyclic $L_\infty$-algebra}} & \shortstack{{non-trivial part of}\\{quantum S-matrix}} \\
 \hline
 \shortstack{{computing minimal model via} \\{homological perturbation theory}} & {Feynman diagrams} \\ [1ex] 
 \hline
\end{tabular}
\end{center}
\caption{\label{tab:Loo}A vocabulary of corresponding notions of $L_\infty$-algebras and quantum field theories.}\vspace{-0.2cm}
\end{center}\end{table}

\paragraph{Example: Yang-Mills theory.}
Consider now the (local) $L_\infty$-algebra $\mathfrak{L}$ from example \eqref{eq:brst}.
We want to consider the standard action functional of a Yang-Mills theory, which is given by
\begin{equation}
    S(A) \;=\; \frac{1}{2}\int_M \langle F_A \,\overset{\wedge}{,}\, \star\! F_A \rangle_\mathfrak{g},
\end{equation}
where $F_A \coloneqq \nabla_{\!A}A = \di A+ [A\,\overset{\wedge}{,}\,A]_\mathfrak{g}$ is the field strength.
By exploiting the existence of the pairing $\langle -\,\overset{\wedge}{,}\,- \rangle_\mathfrak{g}: \Omega^{d-p}(M,\mathfrak{g}) \times \Omega^{p}(M,\mathfrak{g}) \rightarrow \Coo(M)$, we are led to an $L_\infty$-algebra $\mathfrak{Crit}(S)$ whose underlying differential graded vector space is
\begin{equation}
\begin{aligned}
    \mathfrak{Crit}(S)[1] \;=\; \,&\Big( \begin{tikzcd}[row sep={14.5ex,between origins}, column sep= 5ex]
    \Omega^0(M,\mathfrak{g}) \arrow[r, "\di"] & \Omega^1(M,\mathfrak{g}) \arrow[r, "\di\star\di"] & \Omega^{d-1}(M,\mathfrak{g}) \arrow[r, "\di"] & \Omega^{d}(M,\mathfrak{g})
    \end{tikzcd}  \Big)\\
    {\scriptstyle\text{deg}\,=} &\quad \;\;\begin{tikzcd}[row sep={14.5ex,between origins}, column sep= 5ex]
    {\scriptstyle -1} && \quad \;\;{\scriptstyle 0}  && \quad \;\;{\scriptstyle 1}  && \quad \;\;{\scriptstyle 2}
    \end{tikzcd} 
\end{aligned}
\end{equation}
and whose $L_\infty$-algebra structure has only the following non-trivial $L_\infty$-brackets:
\begin{equation}\label{eq:BV-yang-Mills}
    \begin{aligned}
        \ell_1(c) \;&=\; \di c, \qquad &\ell_1(A^+) \;&=\; \di A^+,\\
        \ell_1(A) \;&=\; \di\star\di A, \qquad &\ell_2(c,c^+) \;&=\; [c,c^+]_\mathfrak{g}, \\
        \ell_2(c_1,c_2) \;&=\; [c_1,c_2]_\mathfrak{g}, \qquad &\ell_2(c,A^+) \;&=\; [c,A^+]_\mathfrak{g}, \qquad \\
        \ell_2(c,A) \;&=\; [c,A]_\mathfrak{g},  &\ell_2(A,A^+) \;&=\; [A \,\overset{\wedge}{,}\, A^+]_\mathfrak{g}, \\
    \end{aligned}
\end{equation}\vspace{-0.15cm}
\begin{equation*}
    \begin{aligned}
        \ell_2(A_1,A_2) \;&=\; \di\star[A_1\,\overset{\wedge}{,}\,A_2]_\mathfrak{g} + [A_1 \,\overset{\wedge}{,}\, \star\di A_2]_\mathfrak{g} + [A_2 \,\overset{\wedge}{,}\, \star\di A_1]_\mathfrak{g}, \\
        \ell_3(A_1,A_2,A_3) \;&=\; \big[A_1\,\overset{\wedge}{,}\,\star [A_2\,\overset{\wedge}{,}\,A_3]_\mathfrak{g}\big]_\mathfrak{g} + \big[A_2\,\overset{\wedge}{,}\,\star [A_3\,\overset{\wedge}{,}\,A_1]_\mathfrak{g}\big]_\mathfrak{g} + \big[A_3\,\overset{\wedge}{,}\,\star [A_1\,\overset{\wedge}{,}\,A_2]_\mathfrak{g}\big]_\mathfrak{g}, \\
    \end{aligned}
\end{equation*}
for any $c_k\in\Omega^0(M,\mathfrak{g})$, $A_k\in\Omega^1(M,\mathfrak{g})$, $A^+_k\in\Omega^{d-1}(M,\mathfrak{g})$ and $c^+_k\in\Omega^d(M,\mathfrak{g})$ elements of the underlying graded vector space.
Notice that this is precisely what is known as BV-BRST complex in physics.
Moreover, the classical BV-differential of Yang-Mills theory written above can be presented by a classical BV-action $S_\BV$, so that we have $Q_{\BV}=\{S_\BV,-\}$. Such a BV-action is the following familiar one:
\begin{equation}
    S_\BV(\mathsf{c},\mathsf{A},\mathsf{A}^+,\mathsf{c}^+) \;=\; \int_M \bigg( \underbrace{\frac{1}{2}\langle F_\mathsf{A}, \star F_\mathsf{A} \rangle_\mathfrak{g}}_{S} - \underbrace{\langle \mathsf{A}^+, \nabla_{\!\mathsf{A}}\mathsf{c} \rangle_\mathfrak{g} + \frac{1}{2} \langle \mathsf{c}^+,[\mathsf{c},\mathsf{c}]_\mathfrak{g} \rangle_\mathfrak{g}}_{S_\mathrm{BRST}} \bigg).
\end{equation} \vspace{-0.5cm}

\paragraph{BV-quantization.}
In the $L_\infty$-algebra formulation of BV-theory, one quantizes a field theory by lifting its classical BV-action $S_{\BV}\in\mathcal{O}_{{}_{\!}}\big(T^\vee[-1]E\big)$ to a quantum BV-action $S_{\BV}^\hbar\in\mathcal{O}_{{}_{\!}}\big(T^\vee[-1]E\big)[[\hbar]]$ satisfying the \textit{quantum master equation}
\begin{equation}
    i\hbar\triangle S_{\BV}^\hbar + \frac{1}{2}\{S_{\BV}^\hbar,S_{\BV}^\hbar\} \;=\; 0,
\end{equation}
where $\triangle$ is the BV-Laplacian.
In fact, the introduction of the quantum BV-differential
\begin{equation}
    Q^\hbar_\BV \;\coloneqq\; i\hbar\triangle + \{ S_{\BV}^\hbar ,-\}
\end{equation}
makes the $\mathbb{P}_0$-algebra of observables into a $\mathbb{BD}_0$-algebra (i.e. a Beilinson-Drinfeld algebra), whose structure provides a quantisation of the algebra of observables (see \cite{FactII}).

\subsection{String field theory}

\textit{String field theory }(SFT) is a theoretical framework that reformulates the dynamics of relativistic strings in terms of QFT and, thus, provides their second-quantization. BV-formalism is a natural language to formalize SFT.
For a detailed overview on SFT, see \cite{Maccaferri:2023vns}; for a modern introduction see \cite{Erbin_2021}.

The Hilbert space $\mathfrak{H}$ of a string field theory can be identified with the sub-complex of the BRST-complex of the string on those elements satisfying level matching condition (and some other conditions).
This differential-graded Hilbert space $\mathfrak{H}$ is equipped with an $L_\infty$-algebra structure, where the $n$-bracket $\ell_n$ is given by the $(n+1)$-point function of the closed string.
The inner product of states of the Hilbert space provides the $L_\infty$-algebra $(\mathfrak{H},\ell_n)$ with a cyclic structure $\langle-,-\rangle:\mathfrak{H}\otimes \mathfrak{H} \rightarrow \mathbb{C}$.
The BV-action functional of the bosonic closed string field $\mathit{\Psi}\in\mathfrak{H}$ is given by
\begin{equation}
    S_{\mathrm{BV}}(\mathit{\Psi}) \;=\; \frac{1}{2}\langle\mathit{\Psi},Q\mathit{\Psi}\rangle + \sum_{n>1} \frac{1}{(n+1)!}\langle\mathit{\Psi},\ell_n(\mathit{\Psi},\dots,\mathit{\Psi})\rangle
\end{equation}
where $Q$ is the BRST-differential.
In a similar fashion, one can also write Witten's cubic open string field theory, whose BV-action functional is
\begin{equation}
    S_{\mathrm{BV}}(\mathit{\Psi}) \;=\; \frac{1}{2}\langle\mathit{\Psi},Q\mathit{\Psi}\rangle + \frac{1}{3!}\langle\mathit{\Psi},\ell_2(\mathit{\Psi},\mathit{\Psi})\rangle,
\end{equation}
and which is the only purely cubic covariant string field theory.

\subsection{Derived geometry and shifted Poisson structures}

\paragraph{Derived geometry.}
\textit{Derived algebraic geometry} is a rapidly developing branch within higher geometry that generalizes algebraic geometry to a setting where classical notions of geometry, such as schemes and algebraic stacks, are replaced by their derived counterparts. This generalization allows for a powerful approach to studying geometric objects for the following purposes.
\begin{itemize}
    \item \textit{Intersection theory}: derived geometry provides a natural framework for intersection theory, which is concerned with the intersection of geometric objects.
    \item \textit{Deformation theory}: derived geometry can be used to study the infinitesimal deformation of geometric objects, such as schemes and higher stacks.
\end{itemize}
For an overview of derived symplectic geometry, we redirect to the article \cite{calaque2023derived}. For a detailed overview and discussion in the context of physics, see \cite{Calaque_2021}.

Let $\mathsf{dgcAlg}_\mathbb{k}$ be the simplicial-category of dg-commutative algebras. This can be thought of as the opposite category of \textit{derived affine schemes}, which should be seen as a categorification of the usual notion of affine scheme. So, the $(\infty,1)$-category of \textit{derived algebraic stacks} is defined by $\topos_\mathrm{der} = \mathbf{N}_{hc}([\mathsf{dgcAlg}_\mathbb{k},\mathsf{sSet}]_{\mathsf{proj,loc}}^\circ)$, not too differently from what happens for smooth stacks. 

\begin{figure}[h!]
    \centering 
    \begin{equation*}
    \begin{tikzcd}[remember picture, row sep={14ex,between origins}, column sep={10.5ex,between origins}]
    \text{commutative algebras} \arrow[dd, hook, dashed] \arrow[rrr, "\text{schemes}"]\arrow[ddrrr, "\;\;\;\qquad\text{algebraic stacks}",sloped, start anchor={[xshift=+2ex]}, start anchor={[yshift=+0.8ex]}] &&& \text{sets} \arrow[dd, hook] \\
      \\
    \shortstack{\text{differential-graded}\\\text{commutative algebras}} \arrow[rrr, "\substack{\text{derived}\\\text{algebraic stacks}}", start anchor={[xshift=+0.4ex]}, end anchor={[xshift=-0.4ex]}, " "', sloped, dashed]& && \infty\text{-groupoids}
    \end{tikzcd} \vspace{-0.3cm}
    \end{equation*}
    \caption{A summary family tree of stacks in derived algebraic geometry. }
    \label{fig:derived_geometry}
\end{figure}
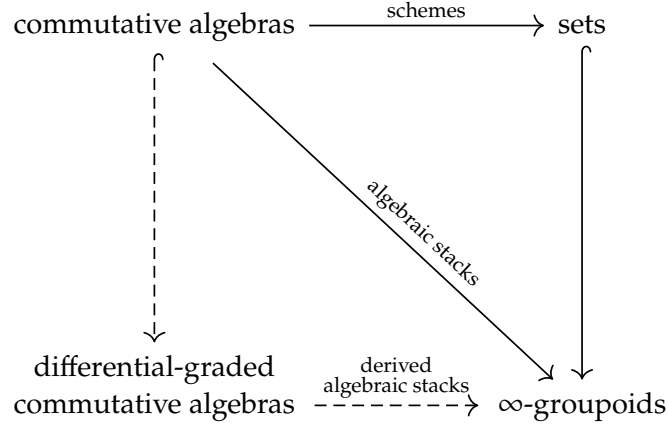

\paragraph{Shifted Poisson structures.}
Let us briefly recall ordinary Poisson structures.
An ordinary Poisson manifold $(X,\pi_2)$ is a manifold $X$ equipped with 
a bivector $\pi_2\in\wedge^2\mathfrak{X}(X)$ such that $[\pi_2,\pi_2]=0$, where $[-,-]$ is the Schouten–Nijenhuis bracket. The $\Coo$-algebra of functions $\Coo(X)$ on a Poisson manifold comes naturally equipped with a Poisson bracket $\pi_2(f,g) \coloneqq \pi_2(\di f\,\overset{\wedge}{,}\,\di g)$. A Poisson structure gives rise to an isomorphism $\pi_2^\flat:T_X^\ast\rightarrow T_X$.
One introduces an ordinary Lie algebroid $\mathfrak{P}(X,\pi_2)$, whose underlying bundle is $T^\ast_X$ and whose Chevalley-Eilenberg dg-algebra is 
\begin{equation}
    \CE(\mathfrak{P}(X,\pi_2)) \;=\; \big(\mathrm{Pol}(X),\; \di_{\pi_2} = [\pi_2,-]),
\end{equation}
where $\mathrm{Pol}(X)= \bigoplus_{n\in\mathbb{N}} \Gamma(X,\wedge^n T_X)$ is the graded algebra of poly-vector fields on $X$ and differential $\di_{\pi_2}\!=\! [\pi_2,-]$.
This provides the archetypal example of the so-called \textit{derived bracket}: in fact, the Poisson algebra can be embedded into the Schouten–Nijenhuis algebra of polyvectors by
\begin{equation}
    \pi_2(f,g) \,=\, [[\pi_2,f],g].
\end{equation}

Let us now give a look at derived version of the story (see e.g. \cite{Pridham:2018, Calaque:2017}). We have a graded commutative algebra of $m$-shifted poly-vectors on a derived scheme $X$ given by
\begin{equation}
    \mathrm{Pol}(X,m) = \bigoplus_{n\in\mathbb{N}}\Gamma(X,\mathrm{Sym}^n\bbT_X[-1-m])
\end{equation}
We can consider its filtration $\mathcal{F}^p\mathrm{Pol}(X,m) = \bigoplus_{n\geq p}\Gamma(X,\mathrm{Sym}^n\bbT_X[-1-m])$.
Similarly to the ordinary case, a $m$\textit{-shifted Poisson structure} is defined by an element $\pi\in\mathcal{F}^2\mathrm{Pol}(X,m)$, that is, a formal sum $\pi = \pi_2 + \pi_3 + \pi_4 + \dots$ where each is of the form $\pi_n\in\Gamma(X,\wedge^n\bbT_X)_{n-2}$ 
such that it satisfies the equation $Q\pi + [\pi,\pi]= 0$, or equivalently the equation
\begin{equation}\label{eq:der_poisson}
    Q\pi_p + \frac{1}{2}\sum_{i+j = p+1} [\pi_i,\pi_j] \;=\; 0,
\end{equation}
for any $p\geq 2$, where $[-,-]$ is the Schouten-Nijenhuis bracket of polyvectors, and $Q$ is the internal differential of $\mathcal{O}(X)$.
Notice that, in analogy with the ordinary case, we can introduce a complex
\begin{equation}
   \big(\mathrm{Pol}(X,m),\; \di_\pi = Q+[\pi,-]\big).
\end{equation}
This construction induces a natural $L_\infty$-algebra structure on the dg-algebra $\mathcal{O}(X)$ of functions on $X$ where the brackets are derivations given by
\begin{equation}
    \pi_p(f_1,f_2\dots,f_p) \,=\, \mathrm{proj}[\cdots[[\pi,f_1],f_2], \cdots f_p],
\end{equation}
where $\mathrm{proj}:\mathrm{Pol}(X)\rightarrow \mathcal{O}(X)$ is the natural projection onto functions on $X$. It is possible to see that the $L_\infty$-bracket structure is implied by defining equations \eqref{eq:der_poisson} of a Poisson structure.
If we denote by $\cdot$ the commutative product of $\mathcal{O}(X)$, seen as a dg-algebra, we have the compatibility condition
\begin{equation}
    \pi_2(g,f\cdot h) \;=\; (-1)^{\left|f\right|\left|g\right|}f\cdot\pi_2(g,h) + \pi_2(g,f)\cdot h
\end{equation}
for any choice of scalars $f,g,h\in\mathcal{O}(X)$. This generalizes the usual compatibility condition between Poisson bracket and commutative product.
The structure $(\mathcal{O}(X), \,\cdot\,, \{\pi_i\}_{i\geq 1})$ is called a \textit{homotopy} $\mathbb{P}_{k+1}$\textit{-algebra}, whose underlying $L_\infty$-algebra\footnote{Notice that the $L_\infty$-algebra $(\mathcal{O}(X), \{\pi_i\}_{i\geq 1})$ generalizes the notion of an ordinary Poisson algebra in a different direction in comparison to the higher Poisson algebra from section \ref{sec:n_plectic}. Informally speaking, the latter is a derived generalization of Poisson algebras, while the latter is higher one.} structure has differential $\pi_1= Q$ and brackets $\{\pi_i\}_{i\geq 2}$.

\subsection{BV-formalism in terms of Factorization Algebras}

For a detailed overview of Factorization Algebras, we redirect to the article \cite{costello2023factorization}. Very roughly, a factorization algebra is a precosheaf of vector spaces on a given manifold $M$ together with a notion of multiplication of sections living on disjoint patches.

\paragraph{Formal Moduli Problems.}
A \textit{formal moduli problem} is, informally speaking, a certain derived stack on dg-Artinian algebras, which are characterized by a maximal differential ideal $\mathfrak{m}_{\mathcal{R}}$ such that $\mathcal{R}/\mathfrak{m}_{\mathcal{R}}\cong \mathbb{R}$. It is possible to show (see \cite{2007arXiv0705.0344P}) that any formal moduli problem is equivalent to one of the form $\mathcal{R}\mapsto\mathrm{MC}(\mathfrak{g}\otimes \mathfrak{m}_{\mathcal{R}})$, where $\mathrm{MC}(\mathfrak{g}\otimes \mathfrak{m}_{\mathcal{R}})$ is the simplicial set of solutions of the Maurer-Cartan equations on $\mathfrak{g}\otimes \mathfrak{m}_{\mathcal{R}}$. The cochain complex of functions on such a formal moduli problem coincides with the (local) Chevalley-Eilenberg cochain $\CE(\mathfrak{g})$.

The first variation of the classical action functional $S\in\CE(\mathfrak{L})$ can be seen as an element
$\delta S\in\CE(\mathfrak{L},\mathfrak{L}^\vee[-1])$ of the local Chevalley-Eilenberg cochain of the BRST $L_\infty$-algebra $\mathfrak{L}$ valued in the $\mathfrak{L}$-module $\mathfrak{L}^\vee[-1]$.
Remarkably, in \cite{FactII} it is shown that the BV $L_\infty$-algebra $\mathfrak{Crit}(S)$ can be geometrically seen as the $L_\infty$-algebra associated to the pointed formal moduli problem which is the \textit{derived critical locus} of the action $S$, i.e. the derived zero locus of $\delta S$.
Thus, the BV $L_\infty$-algebra $\mathfrak{Crit}(S)$ can be obtained from derived intersection theory.
The cochain complex of functions on such a formal moduli problem, which is $\mathrm{Obs}^{\mathrm{cl}} \coloneqq \CE(\mathfrak{Crit}(S))$, can be naturally equipped with the structure of a factorization algebra.

\paragraph{Example: Yang-Mills theory.}
Now, let us now briefly explore a fundamental example of BV-theory in terms of $L_\infty$-algebras and formal moduli problems: Yang-Mills theory.
Let us write explicitly the sets of $0$- and $1$-simplices of the Maurer-Cartan simplicial set $\mathrm{MC}(\mathfrak{Crit}(S)\otimes \mathfrak{m}_{\mathcal{R}})$ for some fixed dg-Artinian algebra $\mathcal{R}$.
The set of $0$-simplices is given by
\begin{equation*}
\begin{aligned}
        \mathrm{MC}(\mathfrak{Crit}(S)\otimes \mathfrak{m}_{\mathcal{R}})_0 \,=\, \left\{ \left. \begin{array}{ll}
         A &\!\!\!\!\in\Omega^1(M,\mathfrak{g})\otimes\mathfrak{m}_{\mathcal{R},0} \\[1.0ex]
        A^+ &\!\!\!\!\in \Omega^{d-1}(M,\mathfrak{g}) \otimes\mathfrak{m}_{\mathcal{R},-1} \\[1.0ex]
        c^+ &\!\!\!\!\in \Omega^{d}(M,\mathfrak{g}) \otimes\mathfrak{m}_{\mathcal{R},-2}
        \end{array}  \;\right|\; \begin{array}{rl} 
        \nabla_{\!A}\star \!F_{\!A} &\!\!\!=\, \di_{\mathcal{R}} A^+ \\[1.0ex]
        \nabla_{\!A}A^+ &\!\!\!=\, \di_{\mathcal{R}} c^+
        \end{array}\right\},
\end{aligned}
\end{equation*}
and the set of $1$-simplices is
\begin{equation*}
\hspace{-0.4cm}\begin{aligned}
        \mathrm{MC}(\mathfrak{Crit}(S)\otimes \mathfrak{m}_{\mathcal{R}})_1 \,=\, \left\{\!\left. \begin{array}{ll}
          c_1 \di t &\!\!\!\!\!\in\Omega^0(M,\mathfrak{g})\otimes\mathfrak{m}_{\mathcal{R},0}\otimes\Omega^1([0,1]) \\[1.0ex]
          A_0 &\!\!\!\!\!\in\Omega^1(M,\mathfrak{g})\otimes\mathfrak{m}_{\mathcal{R},0}\otimes\Omega^0([0,1]) \\[1.0ex]
          A^1\di t &\!\!\!\!\!\in\Omega^1(M,\mathfrak{g})\otimes\mathfrak{m}_{\mathcal{R},-1}\otimes\Omega^1([0,1]) \\[1.0ex]
        A^+_0 &\!\!\!\!\!\in \Omega^{d-1}(M,\mathfrak{g}) \otimes\mathfrak{m}_{\mathcal{R},-1}\otimes\Omega^0([0,1]) \\[1.0ex]
        A^{+}_1\di t &\!\!\!\!\!\in \Omega^{d-1}(M,\mathfrak{g}) \otimes\mathfrak{m}_{\mathcal{R},-2}\otimes\Omega^1([0,1]) \\[1.0ex]
        c^{+}_0 &\!\!\!\!\!\in \Omega^{d}(M,\mathfrak{g}) \otimes\mathfrak{m}_{\mathcal{R},-2}\otimes\Omega^0([0,1]) \\[1.0ex]
        c^{+}_1\di t &\!\!\!\!\!\in \Omega^{d}(M,\mathfrak{g}) \otimes\mathfrak{m}_{\mathcal{R},-3}\otimes\Omega^1([0,1])
        \end{array} \! \right| \begin{array}{rl} 
        \nabla_{\!A_0\!}\star \!F_{\!A_0} &\!\!\!=\, \di_{\mathcal{R}} A^+_0 \\[1.0ex]
        \nabla_{\!A_0}A^+_0 &\!\!\!=\, \di_{\mathcal{R}} c^{+}_0 \\[1.0ex]
        \frac{\di}{\di t}A_0 + \nabla_{\!A_0}c_1 &\!\!\!=\, \di_{\mathcal{R}} A_1  \\[1.0ex]
        \frac{\di}{\di t}A^+_0 +  \nabla_{\!A_0\!}\star\!F_{\!A_1} &\!\!\!\!\!\!+  \\[1.0ex]
        +\, [c_1,A^+_0] &\!\!\!=\, \di_{\mathcal{R}} A^{+}_1 \\[1.0ex]
        \frac{\di}{\di t}c^{+}_0 + \nabla_{\!A_0}A^{+}_1  &\!\!\!\!\!\!+  \\[1.0ex]
        +\, [c_1,c_0^+] &\!\!\!=\, \di_{\mathcal{R}} c^{+}_1
        \end{array} \!\!\right\},
\end{aligned}
\end{equation*}
where $t$ is a coordinate on the unit interval $[0,1]\subset \bbR$.
The elements of this set are $1$-simplices in the sense that each of them links a $0$-simplex
$(A,\, A^+,\, c^+) = (A_0(0),\, A^{+}_0(0),\, c^+_0(0))$ at $t=0$ to the $0$-simplex $(A^{\prime},\, A^{+\prime},\, c^{+\prime}) = (A_0(1),\, A^{+}_0(1),\, c^+_0(1))$ at $t=1$.
And so on for higher simplices. \vspace{0.5cm}

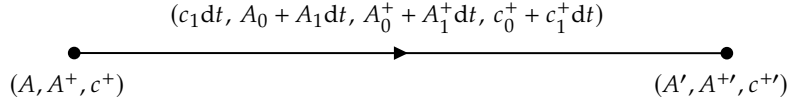
\begin{figure}[h]
    \centering
\tikzset{every picture/.style={line width=0.75pt}} 
\begin{tikzpicture}[x=0.75pt,y=0.75pt,yscale=-1,xscale=1]
\draw    (20.5,49.5) -- (349.67,49.58) ;
\draw [shift={(349.67,49.58)}, rotate = 0.01] [color={rgb, 255:red, 0; green, 0; blue, 0 }  ][fill={rgb, 255:red, 0; green, 0; blue, 0 }  ][line width=0.75]      (0, 0) circle [x radius= 2.68, y radius= 2.68]   ;
\draw [shift={(188.88,49.54)}, rotate = 180.01] [fill={rgb, 255:red, 0; green, 0; blue, 0 }  ][line width=0.08]  [draw opacity=0] (7.14,-3.43) -- (0,0) -- (7.14,3.43) -- cycle    ;
\draw [shift={(20.5,49.5)}, rotate = 0.01] [color={rgb, 255:red, 0; green, 0; blue, 0 }  ][fill={rgb, 255:red, 0; green, 0; blue, 0 }  ][line width=0.75]      (0, 0) circle [x radius= 2.68, y radius= 2.68]   ;
\draw (-13.5,58.4) node [anchor=north west][inner sep=0.75pt]  [font=\footnotesize]  {$( A,A^{+} ,c^{+})$};
\draw (311.5,58.4) node [anchor=north west][inner sep=0.75pt]  [font=\footnotesize]  {$( A',A^{+\prime} ,c^{+\prime})$};
\draw (67.5,22.9) node [anchor=north west][inner sep=0.75pt]  [font=\footnotesize]  {$( c_{1}\mathrm{d} t,\, A_{0} +A_{1}\mathrm{d} t,\, A_{0}^{+} +A_{1}^{+}\mathrm{d} t,\, c_{0}^{+} +c_{1}^{+}\mathrm{d} t)$};
\end{tikzpicture}
    \caption{$0$- and $1$-simplices of $\mathrm{MC}(\mathfrak{Crit}(S)\otimes \mathfrak{m}_{\mathcal{R}})$.}
    \label{fig:bv_fmp}
\end{figure}

\subsection{BV-formalism in terms of Homotopical Algebraic QFT}

Homotopy Algebraic QFT \cite{Benini:2016jfs, Benini:2018oeh, Yau_2019, Benini:2019hoc, Benini:2019uge, Benini:2020skc, Benini:2017fnn, Benini:2020gbr, Bruinsma:2021eis, Anastopoulos:2022cez} provides a differential-graded generalization of the axioms of usual Algebraic QFT, making it a powerful framework adapted to BV-theory and BRST-formalism.

Let $\mathsf{Loc}$ be the category whose objects are oriented and time-oriented globally hyperbolic Lorentzian manifolds and whose morphisms $f:M\rightarrow N$ are orientation and time-orientation preserving isometric embeddings such that the image $f(M)\subset N$ is open and causally convex. Let $\mathsf{dg}\!\ast\!\mathsf{Alg}_\mathbb{C}$ be the category of differential graded unital and associative $\ast$-algebras.
Now, a \textit{homotopy algebraic quantum field theory} is defined as a functor
\begin{equation}
    \mathfrak{A}\,:\,\mathsf{Loc}\,\longrightarrow\, \mathsf{dg}\!\ast\!\mathsf{Alg}_\mathbb{C}
\end{equation}
satisfying the following axioms:
\begin{itemize}
    \item \textit{Einstein causality}: for any pair of morphisms $(f:M\rightarrow N, \,f':M'\rightarrow N)$ in $\mathsf{Loc}$ with causally disjoint images, the morphism in $\mathsf{dg}\!\ast\!\mathsf{Alg}_\mathbb{C}$
    \begin{equation}
        [\mathfrak{A}(f),\mathfrak{A}(f')]:\,\mathfrak{A}(M)\otimes \mathfrak{A}(M') \,\longrightarrow\, \mathfrak{A}(N)
    \end{equation}
     is zero, where $[-,-]$ is the graded commutator of the dg-$\ast$-algebra $\mathfrak{A}(N)$;
    \item \textit{Time-slice}: for any morphism $f:M\rightarrow N$ in $\mathsf{Loc}$ whose image contains a Cauchy surface of $N$, the morphism $\mathfrak{A}(f):\mathfrak{A}(M)\xrightarrow{\,\simeq\,}\mathfrak{A}(N)$ is a weak equivalence in $\mathsf{dg}\!\ast\!\mathsf{Alg}_\mathbb{C}$.
\end{itemize}
Interestingly, in \cite{Benini:2019hoc}, was shown that the quantization of the usual unshifted Poisson structure determines a unique (up to equivalence) homotopy algebraic QFT.

\section{Extended Functorial Quantum Field Theory}

Extended Functorial Field Theory is a higher-categorical generalization of ordinary functorial quantum field theory (FQFT). 
An ordinary FQFT is a monoidal functor $Z : \mathsf{Cob}_d^\sqcup \rightarrow \mathsf{Vect}^\otimes_\mathbb{C}$ from the monoidal category $\mathsf{Cob}_d^\sqcup$ of cobordisms in dimension $d$ to a monoidal category $\mathsf{Vect}^\otimes_\mathbb{C}$ of complex vector spaces.
Thus, by evaluating the functor at a cobordism, one obtains the propagator map $Z(M):Z(\partial_0M)\longrightarrow Z(\partial_1M)$, which relates the Hilbert space at the target to the one at the source of the cobordism; this is strictly related to usual $n$-point functions.

\paragraph{Extended FQFT.}
It is possible to define an $n$-category $n\mathsf{Cob}_d^\sqcup$ of extended cobordisms.
Informally speaking, a ($i+1$)-cobordism between $i$-cobordisms will be a compact oriented ($i+1$)-dimensional smooth manifold (with corners) whose boundary is given by the disjoint union of the target $i$-cobordism and the orientation-reversed source $i$-cobordism.

An \textit{extended functorial quantum field theory} is usually defined as a monoidal $n$-functor
\begin{equation}
    Z \,:\, n\mathsf{Cob}_d^\sqcup \,\longrightarrow\, n\mathsf{Vect}^\otimes_\mathbb{C},
\end{equation}
where $n\mathsf{Vect}^\otimes_\mathbb{C}$ is the $n$-category of $n$-vector spaces over a field, but it can be replaced by a different monoidal $n$-category. 
Extended FQFT has proved to be a powerful language to encode locality in functorial quantum field theories and describe their anomalies (see \cite{Muller:2020phm}).

\paragraph{Chern-Simons as Extended FQFT.} An interesting example of Extended FQFT is given by Chern-Simons theory (see \cite{waldorf2009string}), which can be cast as a functor
$Z_{\xi,k} : 3\mathsf{Cob}_{3}^\sqcup \longrightarrow \mathsf{C}$, for some $G$-bundle $\xi$ on $N$, topological string structure $k\in\mathrm{H}^4(BG,\mathbb{Z})$ and monoidal $3$-category $\mathsf{C}$. 
In \cite[Section 4]{waldorf2009string} the functor $Z_{\xi,k}$ is constructed so that, when evaluated at a $p$-dimensional cobordism $M_{p}$, we have that $Z_{\xi,k}(M_p)$ is the transgression of the Chern-Simons bundle $3$-gerbe $\mathbf{k}(\xi)$ to the mapping space $[M_p,N]$, which is going to be a bundle $(3-p)$-gerbe on $[M_p,N]$.

\section{Higher Kac-Moody algebras}

Higher Kac-Moody algebras \cite{faonte2019higher} are proposed  higher-dimensional generalizations of Kac-Moody algebras, constructed by exploiting derived algebraic geometry. 
In complex dimension $1$, the algebra of $\mathfrak{g}$-valued functions on the punctured affine space is $\mathfrak{g}\otimes\mathcal{O}(\mathbb{A}^1_{\mathbb{C}}-\{0\}) = \mathfrak{g}[z,z^{-1}]$, which allows non-trivial central extensions by $2$-cocycles. These are usual Kac-Moody algebras.
In higher dimension, however, Hartogs' extension theorem tells us that that $\mathfrak{g}\otimes\mathcal{O}(\mathbb{A}^n_{\mathbb{C}}-\{0\}) \cong \mathfrak{g}\otimes\mathcal{O}(\mathbb{A}^n_{\mathbb{C}})$, which implies that there is no interesting central extension. On the other hand, we known that the punctured affine spaces $\mathbb{A}^n_{\mathbb{C}}-\{0\}$ have non-trivial higher cohomology.
The idea, then, is to define higher Kac-Moody algebras by replacing the ordinary algebra $\mathcal{O}(\mathbb{A}^n_{\mathbb{C}}-\{0\})$ with the dg-algebra $\mathbf{R}\Gamma(\mathbb{A}^{n}_{\mathbb{C}}-\{0\},\mathcal{O})$ of derived sections of the structure sheaf, which allows interesting central extensions.
This principle holds also for punctured formal disks $\mathbb{D}_n^\circ$.

\paragraph{Faonte-Hennion-Kapranov higher Kac-Moody algebras.}
Consider a Lie algebra $\mathfrak{g}$. In \cite{faonte2019higher} a \textit{higher current algebra} is defined as the dg-Lie algebra
\begin{equation}
    \mathfrak{g}_n \;\coloneqq\; \mathfrak{g}\otimes\mathbf{R}\Gamma(\mathbb{D}_n^\circ,\mathcal{O}),
\end{equation}
where $\mathbf{R}\Gamma(\mathbb{D}_n^\circ,\mathcal{O})$ is the dg-commutative algebra of derived sections of the structure sheaf $\mathcal{O}$ on the punctured formal disk $\mathbb{D}_n^\circ = \mathrm{Spec}(\mathbb{C} [[z_1,\dots,z_n]] )-\{0\}$.
Then, a \textit{higher Kac-Moody algebra} $\widehat{\mathfrak{g}}_{n,\Theta}$ is defined as the central extension of the higher current algebra $\mathfrak{g}_n$ by an invariant polynomial $\Theta$ of degree $(n+1)$ on $\mathfrak{g}_n$ .

For instance, for $n=1$, one recovers the formal current algebra $\mathfrak{g}_1 = \mathfrak{g}((z))$, whose central extensions $\widehat{\mathfrak{g}}_{1,\Theta}$ are ordinary Kac-Moody algebras.
More generally, for $n>1$, the cohomology of the complex of derived global sections $\mathbf{R}\Gamma(\mathbb{D}_n^\circ,\mathcal{O})$ is
\begin{equation}
    \mathrm{H}^i(\mathbb{D}_n^\circ,\mathcal{O}) \;=\; \begin{cases}\mathbb{C} [[z_1,\dots,z_n]] , & i=0\\z_1^{-1}\cdots z_n^{-1}\mathbb{C}[z_1^{-1},\dots,z_n^{-1}], & i=n-1\\ 0, & \text{otherwise.}\end{cases}
\end{equation} 
In \cite{Alfonsi:2022xiw}, the authors explored a derived version of a Manin triple for higher current algebras.

\paragraph{Relation with higher Kac-Moody factorization algebras.}
Higher Kac-Moody algebras were related by \cite{Gwilliam:2018lpo} to the factorization algebras appearing in BV-theory. 
Consider a complex manifold $X$ with $n\coloneqq\mathrm{dim}_\mathbb{C}(X)$ and equipped with a holomorphic  principal bundle  $P\rightarrow X$ for some structure group $G$. 
Let $Ad(P)$ be the local $L_\infty$-algebra with complex of sections $\Omega_c^{0,\ast}(U,\mathfrak{g}_P)$ on any open set $U\subset X$ and differential given by the $(0,1)$-connection $\bar{\partial}_P$, where $\mathfrak{g}_P \coloneqq P\times_G\mathfrak{g}$ is the adjoint bundle of $P$.
In this setting, we can pick a degree $1$ cocycle $\Theta$ in the local Chevalley-Eilenberg cochains $\mathrm{CE}(Ad(P))$.
Now, in \cite{Gwilliam:2018lpo}, a higher Kac-Moody factorization algebra on $X$ is defined as the twisted universal enveloping factorization algebra $\mathbb{U}_\Theta(Ad(P))$, whose sections are
\begin{equation}
    \mathbb{U}_{\Theta\!}\left(Ad(P)\right)(U) \;=\; \big(\mathrm{Sym}(\Omega_c^{0,\ast}(U,\mathfrak{g}_P)[1]), \; \bar{\partial} + \mathrm{d}_\mathrm{CE} + \Theta \big) 
\end{equation}
on any open set $U\subset X$.
If we denote the radial projection map by $\rho : \mathbb{A}^{n}_{\mathbb{C}} - \{0\} \rightarrow \bbR_{>0}$, a map of factorization algebras on $\bbR_{>0}$ of the following form can be constructed:
\begin{equation}
    \mathbb{U}(\hat{\pi}_{\mathfrak{g},n,\Theta}) \, : \; \mathbb{U}(\Omega^{0,\ast}_c\otimes\hat{\mathfrak{g}}_{n,\Theta}) \; \longrightarrow \; \rho_\ast\mathbb{U}_\Theta(Ad(P)),
\end{equation}
where on the left-hand side we have the enveloping factorization algebra of the \cite{faonte2019higher} higher Kac-Moody algebra $\hat{\mathfrak{g}}_{n,\Theta}$.
Thus, the map above establishes an intriguing relation between derived algebraic geometry and QFT in the language of factorization algebras. In particular, the higher Kac-Moody algebra $\hat{\mathfrak{g}}_{n,\Theta}$ seems to control its corresponding higher Kac-Moody factorization algebra similarly to how an affine Kac-Moody algebra controls its corresponding vertex algebra.

\section{Outlook}

As we have seen in this overview, higher (and derived) geometry provides novel and powerful tools to mathematical physicists. Let us recall some important points.
\begin{itemize}
    \item Smooth stacks provide a natural geometric framework to describe configuration spaces of gauge theories;
    \item Higher gauge theories naturally emerge in string theory. For example, the gauge sector of Type II and Heterotic string theory can be naturally formalized as higher gauge theories. 
    \item Higher geometric quantization allows to non-perturbatively prequantize quantum field theories, including bosonic strings and Chern-Simons theory. 
    Hypothesis H goes beyond this, proposing a way to charge-quantize M-theory.
    \item Derived geometry formalizes BV-theory, which gives us a framework to place our field theories on-shell, reproduce perturbation theory and compute scattering amplitudes.
\end{itemize}

\paragraph{The puzzle of quantization.}
In this article we encountered two distinct notions of quantization.
\begin{itemize}
    \item Higher geometric quantization, despite its beauty, is still affected by the important problem of not coming equipped with a good notion of polarisation of the prequantum bundle, implying that we do not have a fully fledged Hilbert space.
Nonetheless, higher geometric quantisation reminds us of the crucial lesson that quantization is ultimately a global-geometric process. 
    \item In contrast, BV-theory works very well in reproducing scattering amplitudes, but it is intrinsically perturbative. Geometrically, this is because field theories are deformation-quantized in a series expansion around a fixed solution. 
\end{itemize}
In some sense, features and limitations of the two frameworks appear complementary.
Remarkably, in derived algebraic-geometry, the two quantizations have been related by \cite{safronov2020shifted}.
This suggests the existence of an intriguing connection between geometric quantization and BV-quantization. It is possible to speculate that they could be unified in a single higher derived quantization. 

\section*{Acknowledgments}

The author gratefully acknowledges the financial support of the Leverhulme Trust, Research Project Grant number RPG-2021-092. 


\setlength{\baselineskip}{0pt}
\renewcommand*{\bibfont}{\scriptsize}

{\renewcommand*\MakeUppercase[1]{#1}%
\printbibliography[heading=bibintoc,title={\bibtitle}]}

\end{document}